\documentclass[a4paper]{article}

\usepackage[pages=all, color=black, position={current page.south}, placement=bottom, scale=1, opacity=1, vshift=5mm]{background}
\backgroundsetup{contents={}}

\usepackage{natbib,amsmath}

\usepackage[margin=1in]{geometry} 

\title{Dynamic stall of a hydrofoil with tubercles\\in surface gravity waves}

\author{Guillaume Ricard\textsuperscript{1,2}, Gunnar Jacobi\textsuperscript{1}, Daniele Fiscaletti\textsuperscript{1}, Abel-John Buchner\textsuperscript{2}\\ \\
\small \textsuperscript{1}Department of Maritime \& Transport Technology, Faculty of Mechanical Engineering,\\
\small Delft University of Technology, Delft, The Netherlands\\
\small \textsuperscript{2}Department of Process \& Energy, Faculty of Mechanical Engineering,\\
\small Delft University of Technology, Delft, The Netherlands\\ \\
\small Corresponding author: a.j.buchner@tudelft.nl}

\date{}

\begin{document}
\maketitle

\begin{abstract} 
The interaction of an object with an unsteady flow is non-trivial and is still far from being fully understood. When an airfoil or hydrofoil, for example, undergoes time-dependent motion, nonlinear flow phenomena such as dynamic stall can emerge. The present work experimentally investigates the interaction between a hydrofoil and surface gravity waves. The waves impose periodic fluctuations of the velocity magnitude and orientation, causing a steadily translating hydrofoil to be susceptible to dynamic stall at large wave forcing amplitudes. Simultaneous measurement of both the forces acting on the hydrofoil and the flow around it by means of particle image velocimetry (PIV) are performed, to properly characterise the hydrofoil-wave interaction. In an attempt at alleviating the impact of the flow unsteadiness via passive flow control, a bio-inspired tubercle geometry is applied along the hydrofoil leading edge. This geometry is known to delay stall in steady cases but has scarcely been studied in unsteady flow conditions. The vortex structures associated with dynamic stall are identified, and their trajectories, dimension, and strength characterised. This analysis is performed for both straight- and tubercled-leading-edge geometries, with tubercles found to qualitatively modify the flow behaviour during dynamic stall. Contrary to previous studies, direct measurements of lift do not evidence any strong modification by tubercles. Drag-driven horizontal force fluctuations, however, which have not previously been measured in this context, are found to be strongly attenuated. This decrease is quantified, and a physical model based on the flow observations is finally proposed.
\end{abstract}

\section{Introduction}\label{intro_section}
A hydrofoil, or aerofoil, in a steady flow is a configuration with obvious industrial relevance, for example in aerospace or wind energy applications. It therefore represents one of the most widely studied flows over the last century~\citep{abbott1959doenhoff,clancy1975aerodynamics}. In comparison, studies on the impact of an unsteady flow are relatively new and evidence different flow behaviours emerging when a foil undergoes strong angle-of-attack variation. In the unsteady aerodynamics literature, such variation is often stylised as pitching, heaving, or surging motions, or a combination of these~\citep{eldredge2019leading}. Such variations, when they are of sufficient magnitude and rate, lead to dynamic stall, wherein flow separates from the foil leading edge and a stall vortex is formed. Lift and drag are in this case temporarily enhanced over the time scale of dynamic stall vortex formation and shedding
~\citep{carr1988progress,ekaterinaris1998computational,mulleners2012onset,choudhry2014insight}. Dynamic stall-related phenomena are important to several fields of aerodynamics, including insect flight, wherein organisms leverage high instantaneous force generation for manoeuvrability~\citep{dickinson1999wing,ellington1996leading}, wind energy where oscillatory loading is relevant to fatigue~\citep{simao2009visualization,buchner2018stall}, and helicopter rotor aerodynamics where it limits the forward flight speed~\citep{carr1977analysis}.

Instead of foil motion, dynamic stall can instead be triggered by unsteady variation in the imposed flow, such as due to wind gusts or wave action. Surface gravity waves induce an orbital motion of the fluid beneath them, characterised by strong and successive variations of the flow direction and magnitude. To the best of our knowledge, a characterisation of the dynamic stall regime generated by the action of waves on a horizontal foil has not previously been reported in the scientific literature. 
The combination of periodic variations in both flow magnitude and direction makes this a fundamentally different case than any of the combinations of the aforementioned pitch/heave/surge stylisations of unsteady motion which have been previously investigated in the literature. 
That is not to say that no literature exists on the interaction of a body with waves. Indeed, this is a key driver of hydrodynamics research in offshore engineering. 
Forces generated by waves acting on an immersed cylinder have, for example, been extensively characterised and found to be driven by drag and inertia~\citep{sarpkaya1977line,chaplin1984nonlinear,chaplin1997large,venugopal2009drag,bai2017study}. Forces acting on an immersed vessel under linear waves have also been described numerically, showing the importance of surge and heave forces as well as of pitch moment~\citep{malik2013research}. There also exists literature on the effect of waves on tidal (hydrokinetic) turbines, where it has been experimentally~\citep{galloway2014quantifying} and numerically~\citep{scarlett2020unsteady} shown that the wave frequency and the wave amplitude drive the force fluctuations that act on a turbine. The average axial load and torque experienced by a tidal turbine are similar with or without small amplitude waves, but wave-induced oscillatory force fluctuations can be harmful as they can increase the fatigue of the turbine structure~\citep{draycott2019experimental,guo2018surface}. Understanding to what extent surface waves impact a turbine is therefore essential for predicting turbine performance and modelling fatigue, and for optimizing turbine design and operation to avoid the negative aspects of unsteady hydrodynamic forcing. The unsteadiness could even represent an opportunity for performance gain, given the right strategy~\citep{wei2015experimental}. This motivates an investigation into the unsteady mechanisms governing the effect of wave-induced flows on submerged objects. Here, we consider the problem of a horizontally-mounted hydrofoil under wave action. This represents a model system for understanding wave-induced dynamic stall, while avoiding the complexity associated with more complicated geometries or kinematics, such as those associated with hydrokinetic turbines. 
Interaction between a hydrofoil and solitary internal waves has previously been studied by imaging the flow and measuring how forces acting on the foil vary~\citep{zou2023experimental}, but this work did not include data in a stall regime. The present work focuses first on how dynamic stall can be generated on a steadily moving horizontal hydrofoil by the orbital flow induced by waves. 
This is done experimentally by measuring the forces acting on the hydrofoil using a six-component force balance, and by simultaneously visualizing the flow, driven by waves, around the hydrofoil via particle image velocimetry (PIV). These two measurements, simultaneously performed, enable a direct link to be drawn between the flow behaviour and the resulting force generation. To this end, detailed analysis of the generation and evolution of vortex structures in the flow will follow.

As waves could be expected to cause potentially large magnitude temporal force variation, including as a result of dynamic stall, their action could damage an immersed hydrofoil, for example one fitted to a hydrokinetic turbine, via increased fatigue~\citep{draycott2019experimental,guo2018surface}. This motivates consideration of methods for the mitigation of dynamic stall effects. Several previous attempts have been made in this regard. Self-supplying air jet vortex generators~\citep{krzysiak2013improvement} or adaptive leading-edge geometry~\citep{kerho2007adaptive}, for example, have been employed to effectuate flow control in an active way. Passive flow control has also been attempted, involving fixed modification of the leading-edge geometry. Spanwise-periodic modulation of the leading edge geometry, for example, is commonly known as ``tubercles''~\citep{fish1995hydrodynamic}. This class of leading edge geometry modification is inspired by the pectoral fins of the humpback whale, \textit{Megaptera novaeangliae}, where a series of leading-edge protuberances has been shown to improve their swimming capacity by enhancing the animal manoeuvrability via delayed stall, and some resulting combination of increased lift or drag force generation~\citep{miklosovic2004leading}. The fluid dynamic effect of this geometry has been characterised on foils in steady flow both experimentally~\citep{johari2007effects,wei2015experimental,shi2016numerical,fan2022numerical} and numerically~\citep{fan2022numerical,shi2016numerical}. Tubercles have been found to act as vortex generators that create streamwise vortices which enhance surface-normal momentum transfer, thus energising the boundary layer over the foil's suction surface
~\citep{wei2015experimental,shi2016numerical,fan2022numerical}. This modification in the flow delays flow separation (stall) and so increases the generation of lift force at large angle of attack~\citep{miklosovic2004leading,johari2007effects}, while also potentially affecting the incidence of cavitation at high Reynolds number
~\citep{li2023hydrofoil}. 
Tubercles have also been used on rotating turbines, in which case the boundary layer was also found to remain attached to the blades longer when compared to a straight leading edge geometry. This resulted in an observed increase in harvested power, as long as the tip speed-ratio stays small enough~\citep{shi2017detailed,fan2023effect}. The use of tubercles 
in unsteady flow applications has, on the other hand, been quite scarcely investigated. 
Two recent works address this problem, focusing on the effects of tubercles on dynamic stall of a pitching airfoil~\citep{hrynuk2020effects,valls2025dynamic}. The two studies are not in the exact same range of pitch rate, but both observed similar qualitative dynamic stall behaviour, showing that stall occurs first between tubercles before progressing to other planes. In~\citet{hrynuk2020effects}, with the use of tubercles the lift is hypothesized to be enhanced by the presence of a dynamic stall vortex that is closer to the hydrofoil surface and with greater circulation, whereas, in the regime studied in~\citet{valls2025dynamic}, tubercles are found to mitigate dynamic stall and to decrease the lift fluctuations. These differences demonstrate the lack of coherent understanding in the current state-of-art of the effect of tubercles on dynamic stall. The present study applies the tubercle geometry to a hydrofoil in a wave-induced oscillatory flow. The wave flow being a unique method of generating dynamic stall, the application of tubercles will act as a passive flow technique that could increase lift forces~\citep{hrynuk2020effects} and/or reduce forces fluctuations~\citep{valls2025dynamic}, both results being advantageous regarding the use of a potential hydrokinetic turbine. The novelty of the current work is not only brought by the use of waves to trigger dynamic stall, but also by the simultaneous use of force measurements and of flow visualization, which was not done in the previous works on unsteady effects and tubercles in~\citet{hrynuk2020effects} and \citet{valls2025dynamic}.

The present study has two objectives. First, the goal is to quantify the flow field driven by waves around a hydrofoil, and the fluctuating forces generated by this flow. This quantification is performed across a range of wave conditions including both attached flow and dynamic stall. Then, as dynamic stall is triggered by some wave conditions, the implementation of tubercles is studied as a means to modify the flow and, consequently, the forces. In the second section of the present paper, the flow field induced by waves is detailed and the expected results are theoretically discussed. Then, in the third section, the experimental approach based on the use of a 142 meter long water tank facility, is presented. In the fourth section, results regarding the characterisation of the hydrofoil-wave interaction are detailed. The horizontal and vertical force variations are shown, and the flow field is quantified by means of particle image velocimetry (PIV). Subsequently, the implementation of tubercles on the hydrofoil leading-edge and the hydrodynamic effects of these tubercles' presence are discussed. The last part analyses the vorticity generation and behaviour more in-depth, comparing the two hydrofoils with differing leading edge geometry.

\section{Theoretical background}\label{theory_section}
The present study considers the superposition of a steady horizontal flow of velocity $\textbf{u}_0=u_0 \textbf{e}_x$ with an unsteady flow $\textbf{u'}=u_x'(t,x,z)\textbf{e}_x+u_z'(t,x,z)\textbf{e}_z$ imposed by the orbital velocity of gravity surface waves. The wave dispersion relation of the gravity surface waves in the deep-water approximation, accounting for a Doppler shift, reads

\begin{equation}
    (\omega_r-u_0k)^2=gk,
    \label{relat_disp}
\end{equation}
with $\omega_r=2\pi f$ the angular velocity and $f$ the wave frequency perceived by the hydrofoil, $k=2\pi/\lambda$ the wave number, and $\lambda$ the wave length. $g=9.81$~m/s$^2$ is the gravitational acceleration. The wave amplitude, $A$, is quantified by the wave steepness $\epsilon=2A/\lambda$. The orbital motion of radius $\mathcal{R}(z)=Ae^{\frac{2\pi z}{\lambda}}$ and its tangential velocity $u'(z)=\mathcal{R}(z)\omega$, with $\omega=\sqrt{gk}$ given by Eq.~\eqref{relat_disp} for $u_0=0$~m/s, decrease exponentially with depth $z$ (here negative).

\begin{figure}[h!]
    \centering
    \includegraphics[width=0.8\linewidth]{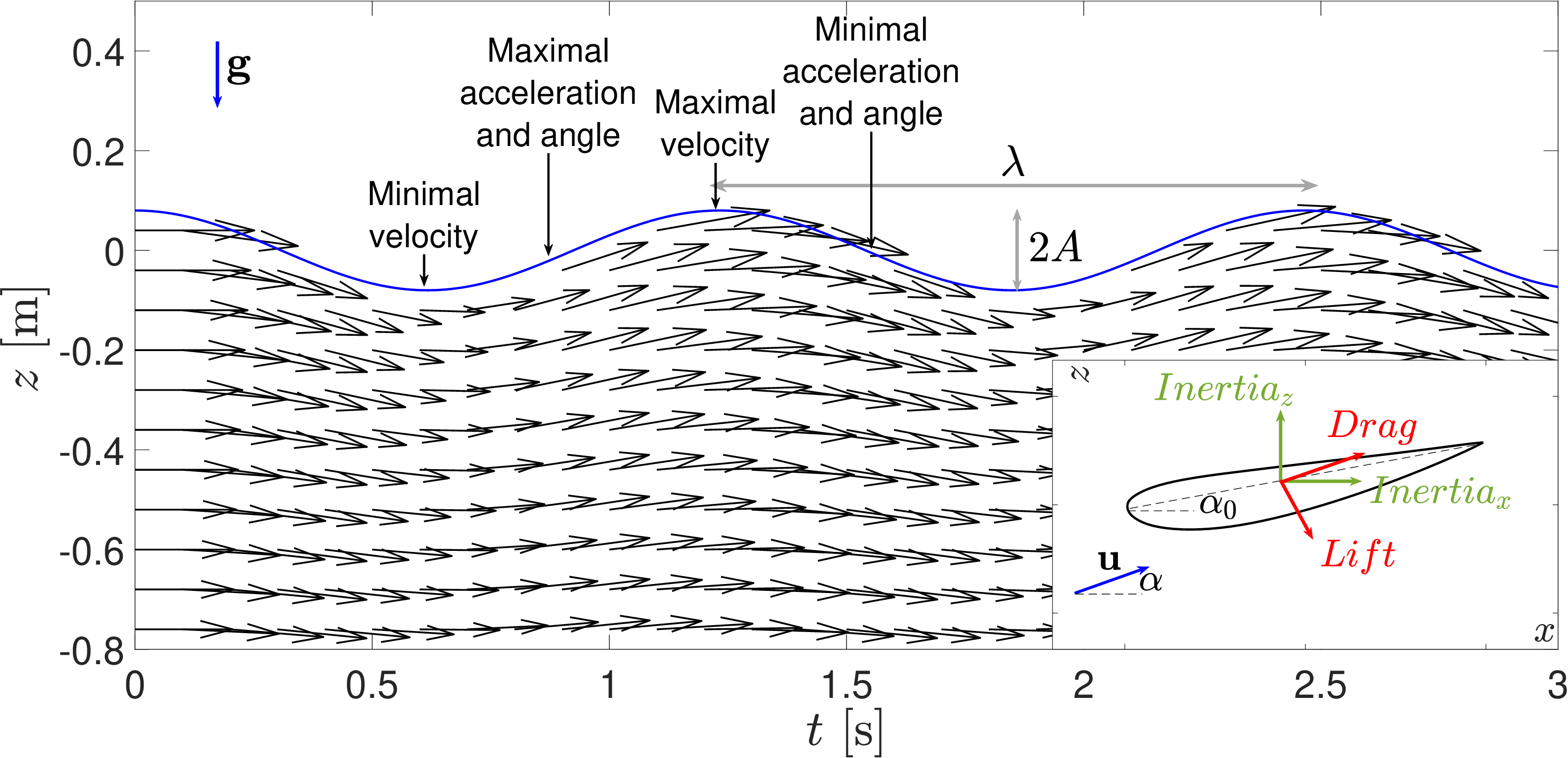}
    \caption{Theoretical velocity field expected at a fixed abscissa $x$, under surface waves calculated using potential flow theory with $u_0=0.75$ m/s, $\lambda=4$~m and $\epsilon=0.04$. The locations of the extrema of velocity magnitude, acceleration and angle of attack are shown. Inset: schema of the hydrofoil used with the direction of the different forces indicated.}
    \label{lin_field}
\end{figure}

The total velocity field $\textbf{u}_0+\textbf{u'}$ in the flow is estimated using linear potential flow theory and is represented in Fig.~\ref{lin_field} for a selected test condition of $u_0=0.75$ m/s, $\lambda=4$~m, and $\epsilon=0.04$. The presence of a large constant velocity component $u_0$ dominates the wave-induced orbital motions, with an immersed object experiencing successive variations of relative velocity magnitude and direction about the mean flow $\textbf{u}_0$. The relative velocity magnitude is maximum at the wave peaks and minimum at their trough locations, implying a maximal and minimal acceleration on the rising and descending wave slopes, respectively. The direction of the local velocity vector that will induce an angle of attack between the flow and the hydrofoil is estimated as

\begin{equation}
    \phantom{.}\tan{(\alpha)}=\frac{u_z}{u_x}=\frac{\mathcal{R}\omega\sin(kx-\omega_r t)}{\mathcal{R}\omega\cos(kx-\omega_r t)+u_0},
    \label{tanalph_eq}
\end{equation}
where $u_x=u_0+u_x'$ and $u_z=u_z'$ are respectively the horizontal and vertical velocity components of the flow seen by the hydrofoil. The amplitude of the flow velocity fluctuation in the vertical direction $z$, is hence given by $\mathcal{R}\tan(\alpha)$. Note that the flow field shown here is calculated through linear approximation, waves with large amplitude (typically $\epsilon\geq0.04$) will exhibit nonlinearities not taken into account in the present estimation.

The expected horizontal and vertical forces $F_x$ and $F_z$ experienced by the immersed hydrofoil follow from the Morison equation~\citep{morison1950force,chaplin1997large} accounting for 
drag, lift, and inertia effects. They are constructed as

\begin{equation}
    \begin{split}
        F_x &=\frac{1}{2}C_d(\beta) \rho csu^2\cos(\alpha)+\frac{1}{2}C_l(\beta)  \rho csu^2\sin(\alpha)+C_{mx}(\beta)  \rho V \frac{\partial u}{\partial t}+F_f,\\
        F_z &=-\frac{1}{2}C_l(\beta)  \rho csu^2\cos(\alpha)+\frac{1}{2}C_d(\beta) \rho csu^2\sin(\alpha)-C_{mz}(\beta)  \rho V \frac{\partial u}{\partial t},\\
    \end{split}
    \label{Force_eq}
\end{equation}
where $\rho$ is the liquid density, $V$ the hydrofoil volume, and $\beta=\alpha_0-\alpha$ the angle of attack between the flow and the hydrofoil. $\alpha_0$ is the static pitch angle of the hydrofoil. $C_d$, $C_l$, $C_{mx}$ and $C_{mz}$ are respectively the drag, lift, horizontal and vertical inertia coefficients. A constant force $F_f$ is added to the horizontal component to account for the friction of the end plates that are used to mount the hydrofoil in the water and enforce quasi-two dimensionality in the flow. The directions of the different forces acting on the hydrofoil are represented in the inset of Fig.~\ref{lin_field}. Note that because of experimental constraints, detailed later, the hydrofoil is mounted upside down, implying a negative lift and the use of $\beta=\alpha_0-\alpha$ as angle of attack. The drag and the lift are defined respectively to be aligned and perpendicular with the local flow vector, whereas the inertia forces are defined along the global $x$ and $z$ coordinates. The force coefficients depend on the angle of attack between the hydrofoil and the local flow velocity, which also varies with time. As indicated by Eq.~\eqref{Force_eq}, the flow velocity, acceleration, and angle of attack will all affect the total force generation. Competition between these external parameters, them not being in phase, drives variation in $F_x$ and $F_z$.

\section{Experimental setup}\label{setup_section}
\begin{figure}[h!]
    \centering
    \includegraphics[width=1\linewidth]{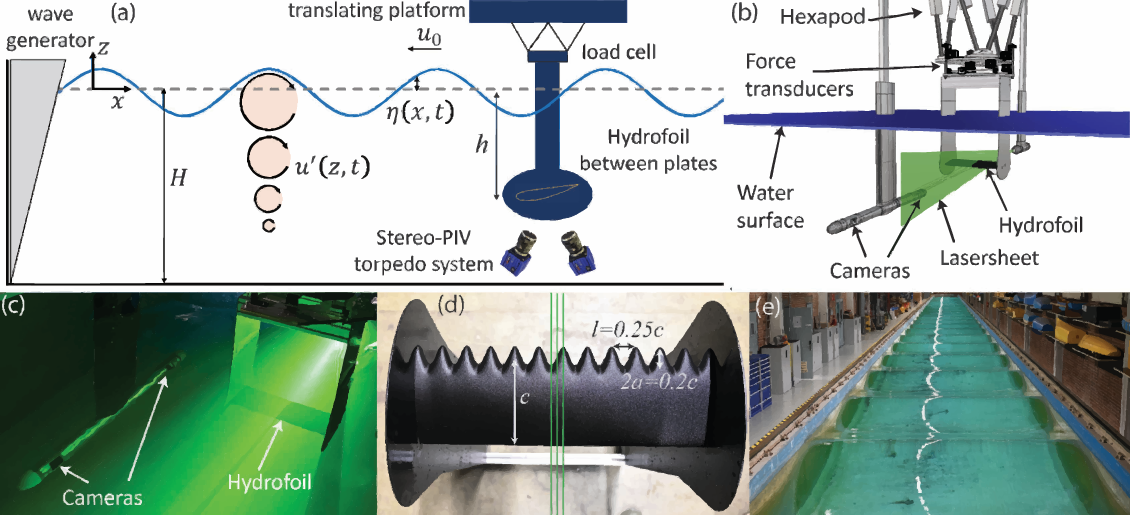}
    \caption{(a) Experimental setup, implemented in a 142~m long towing tank of $H=2.3$~m water depth. A hydrofoil (yellow outline) is immersed in water between two end plates and attached to a translating platform that moves along the tank at a speed $u_0$. Waves are generated by a wavemaker, implying an orbital flow $u'$. (b) Model of the setup in which the PIV system (cameras and laser sheet) is fully represented. The location of the force transducers is indicated. The hexapod to which the model is mounted does not impose any motion in the currently reported experiment. (c) Photograph of the immersed model, with the PIV system in operation. (d) Photograph of hydrofoil with tubercles on its leading-edges, the three planes in which PIV measurements are taken are represented by green vertical lines. (e) Photograph of the towing tank with waves of wave length $\lambda=4$~m and steepness $\epsilon=0.04$.}
    \label{setup}
\end{figure}

Experiments were performed in a 142~m long towing tank of depth $H=2.3$~m at Delft University of Technology. The experimental setup is shown in Fig.~\ref{setup}(a), where the hydrofoil shape is represented as a yellow outline. The hydrofoil is mounted with a baseline angle of attack of $\alpha_0=10^\circ$ to a translating platform that moves along the tank at a constant speed $u_0=0.75$~m/s. This constant speed imposes a steady constant flow on the hydrofoil. The hydrofoil (NACA4415), also represented in Fig.~\ref{setup}(b) and shown in (c), of chord length $c=0.16$~m~$\ll\lambda$ [scaling not respected in Fig.~\ref{setup}(a)] and span $s=0.48$~m, is immersed in water at a constant distance from the mean free surface height of $h=0.4$~m. The immersion depth being 2.5 times larger than the chord length, free-surface effects are negligible and do not affect hydrodynamic forces~\citep{pernod2023free}. End plates are used to attenuate the transversal flow component, mimicking a 2D flow in the $(x,z)$ plane. Two different models are tested, first, one with a uniform leading edge as in Fig.~\ref{setup}(b-c), and then a model fitted with tubercles, shown in Fig.~\ref{setup}(d). The tubercle geometry used is spanwise sinusoidal, with amplitude $0.1 c=16$~mm and wave length $0.25 c=40$~mm, 
following the geometry described in~\citet{wei2015experimental} and used in~\citet{hrynuk2020effects} and~\citet{valls2025dynamic}.

At one extremity of the tank, a wave generator creates sinusoidal gravity waves of chosen wave length $\lambda\in[2,6]$~m and amplitude, i.e., of chosen steepness $\epsilon\in[0.02, 0.06]$. At the other extremity of the tank, a parabolic beach is used to dissipate the waves, thus avoiding most reflections. An example of typical waves generated, with $\lambda=4$~m and $\epsilon=0.04$, is shown in Fig.~\ref{setup}(e). The wave surface elevation $\eta(t)$ is measured using an acoustic sensor, which is traversed with the hydrofoil, at an acquisition rate of $1$~kHz.  
The waves, as observed in a reference frame co-translating with the hydrofoil, impose an orbital velocity field $u'(z,t)$ that makes the total flow field $u=u_0+u'$ experienced by the hydrofoil unsteady (see Fig.~\ref{lin_field}). In order to quantify the flow variation, a stereoscopic particle image velocimetry (PIV) system is suspended from the translating platform. The PIV setup, shown in Fig.~\ref{setup}(b) and in operation in Fig.~\ref{setup}(c), includes two cameras mounted in a torpedo structure to the side of the hydrofoil, and a laser sheet projected from the downstream direction in the $(x,z)$~plane. Because the end plates obscure the flow above the hydrofoil, measurements of the flow are made from below; the hydrofoil was mounted upside down for this purpose. Frame pairs, separated by 1~ms, are recorded by the two cameras at an acquisition rate of 45~Hz. Velocity fields are computed using DaVis 10.2.1.90519, via cross-correlation using an initial interrogation window of $96\times96$ pixels and a final of $64\times64$ pixels, with 50\% overlap, yielding a final vector spacing of $3.4$~mm, or approximately 50 vectors per chord length of the hydrofoil. Although both in-plane and spanwise velocity components are measured, we focus in the present paper on in-plane phenomena, while the $y$ direction component is left for future treatment. The horizontal, $F_x$, and vertical, $F_z$, forces acting on the hydrofoil are obtained from a six-component force balance at an acquisition rate of $1$~kHz. A low-pass filter, with cut-off frequency 5~Hz, is used to remove noise that mainly originates from high-frequency vibrations of the towing platform. Measurements are performed on both hydrofoil models for a range of different wave conditions. 

The parameters chosen here result in a Reynolds number $Re=u_0c/\nu\approx1.2\times10^5$, with $\nu=10^{-6}$~m$^2$/s the water kinematic viscosity. To quantify the unsteady effects of waves on the hydrofoil, the Keulegan–Carpenter number, $KC=\Delta u_x/(fc)\in[0.15,3.97]$, is used, with $\Delta u_x$ the amplitude of velocity variation and $f$ the wave encounter frequency of the hydrofoil. The values of $KC$ in this experiment indicate that inertia forces (i.e.\ added mass) dominate or are of similar order to steady drag forces~\citep{keulegan1958forces}. The rate of change of angle of attack can be expressed as a pitch rate~\citep{hrynuk2020effects,valls2025dynamic}, non-dimensionalised as $\Omega^*=\Delta\alpha fc/U_0\in[0.01,0.06]$, with $\Delta\alpha$ the amplitude of variation of the angle of attack. Unsteady effects associated with vortex formation are expected to emerge from $\Omega^*>0.005$, whereas from $\Omega^*\geq0.05$ a quasi-static description is no longer valid and a dynamic stall regime dominated by unsteady aerodynamic effects is achieved~\citep{choudhry2014insight}. Regarding the experimental parameters chosen here, both regimes should be observable. The various experimental conditions used in the present work are summarized in table~\ref{tab_exp}. The maximum and minimum phase-averaged values of the imposed velocity components $u_x$ and $u_z$ and of the flow orientation $\alpha$ are estimated via PIV in an area upstream of the leading edge at one chord length below the hydrofoil. This location is considered to represent the external flow field, as it is far enough from the hydrofoil so as to barely be influenced by the hydrofoil's presence. 
The case without waves (first row of table~\ref{tab_exp}) shows this to not be precisely true for the horizontal velocity component $u_x=0.8$~m/s, which is slightly larger than the constant value $u_0=0.75$~m/s expected for this case. The value of $u_x$ is thus slightly overestimated in our measurements. As the hydrofoil is upside down with a baseline pitch angle $\alpha_0=10^\circ$, the angle of attack is given by $\beta=\alpha_0-\alpha$. Note that the amplitude of variation of the angle of attack is here at most $25^\circ$, which is smaller than the values of $30^\circ$ and $50^\circ$ used in \citet{valls2025dynamic} and \citet{hrynuk2020effects}, respectively. Note also that the case $\lambda=6$~m, $\epsilon=0.06$ is not tested because the large wave amplitude led to indications of wave breaking phenomena in our experimental facility.

\begin{table}
  \begin{center}
\def~{\hphantom{0}}
  \begin{tabular}{ccccccccccc}
      $\lambda$& $\epsilon$& $\alpha_{min}$& $\alpha_{max}$& $u_{x,min}$& $u_{x,max}$& $u_{z,min}$& $u_{z,max}$& $KC$& $\Omega^*$\\[3pt]
       0 m& 0& 0$^\circ$& 0$^\circ$& 0.80 m/s& 0.80 m/s& 0 m/s& 0 m/s& 0& 0\\
       2 m& 0.02& -1.4$^\circ$& 1.4$^\circ$& 0.78 m/s& 0.81 m/s& -0.02 m/s& 0.02 m/s& 0.15& 0.01\\
       2 m& 0.04& -2.3$^\circ$& 2.5$^\circ$& 0.77 m/s& 0.83 m/s& -0.03 m/s& 0.04 m/s& 0.30& 0.02\\
       2 m& 0.06& -3.0$^\circ$& 3.5$^\circ$& 0.75 m/s& 0.85 m/s& -0.04 m/s& 0.05 m/s& 0.50& 0.03\\
       4 m& 0.02& -3.5$^\circ$& 3.9$^\circ$& 0.74 m/s& 0.86 m/s& -0.05 m/s& 0.05 m/s& 0.93& 0.02\\
       4 m& 0.04& -6.7$^\circ$& 7.4$^\circ$& 0.68 m/s& 0.91 m/s& -0.10 m/s& 0.10 m/s& 1.77& 0.04\\
       4 m& 0.06& -10.1$^\circ$& 10.9$^\circ$& 0.64 m/s& 0.96 m/s& -0.15 m/s& 0.14 m/s& 2.47& 0.06\\
       6 m& 0.02& -5.5$^\circ$& 6.1$^\circ$& 0.69 m/s& 0.90 m/s& -0.08 m/s& 0.08 m/s& 2.08& 0.03\\
       6 m& 0.04& -12.1$^\circ$& 13.4$^\circ$& 0.60 m/s& 1.00 m/s& -0.17 m/s& 0.18 m/s& 3.97& 0.06\\
  \end{tabular}
  \caption{Experimental conditions.}
  \label{tab_exp}
  \end{center}
\end{table}

\section{Effects of waves on a regular hydrofoil}\label{regfoil_section}
\begin{figure}[h!]
    \centering
    \includegraphics[width=1\linewidth]{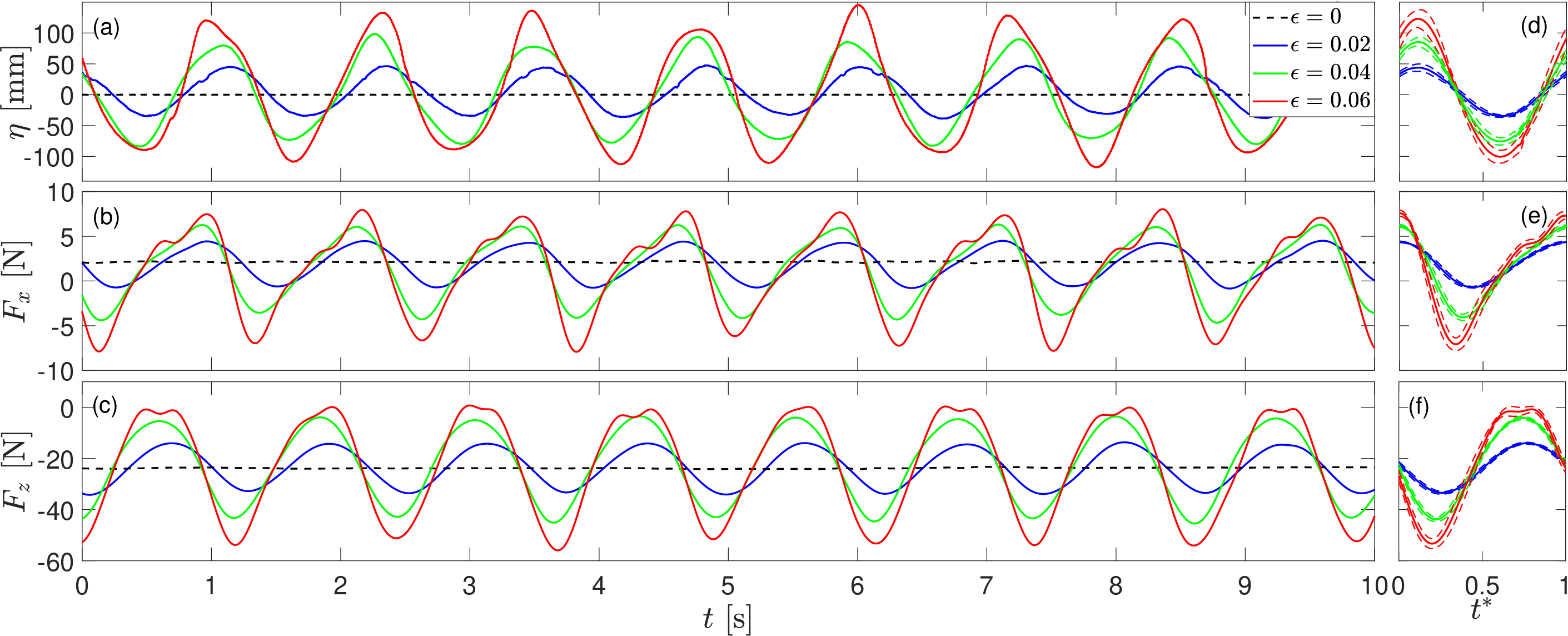}
    \caption{Time history of the surface elevation $\eta(t)$ (a), the horizontal force component $F_x$ (b), and the vertical force component $F_z$ (c), for $\lambda=4$~m and various wave amplitudes: $\epsilon=0$ (black dashed lines), 0.02 (blue solid lines), 0.04 (green solid lines) and 0.06 (red solid lines). (d,e,f) Phase-averaged data over 40 wave periods of the same. Dashed lines correspond to the standard deviation of the average.}
    \label{etaFxFzNotub}
\end{figure}

We first focus on how waves interact with a hydrofoil with a straight-leading-edge geometry. Simultaneous time histories (a) of the surface elevation $\eta(t)$, and (b) of the horizontal and (c) vertical forces $F_x(t)$ and $F_z(t)$ are plotted in Fig.~\ref{etaFxFzNotub}, for $\lambda=4$~m and various wave amplitudes, $\epsilon=0$, i.e.\ no waves, (black dashed lines), 0.02 (blue solid lines), 0.04 (green solid lines) and 0.06 (red solid lines). Phase-averaged data over 40 wave periods of the same quantities are plotted in Fig.~\ref{etaFxFzNotub}(d-f) (solid lines) with their standard deviation (dashed lines), where the time $t^*$ is non-dimensionalised by the wave period $T=2\pi/\omega_r$. Time histories of other incident wave lengths are plotted in the Supplemental Material in Fig.~S1 and S2, and evidence similar results. 

In the case $\epsilon=0$, no wave is generated so the angle of attack is only given by the baseline pitch angle $\beta=\alpha_0=10^\circ$. The quantities $\eta=0$~mm, $F_x\approx2.1$~N, and $F_z\approx-23.8$~N remain constant over time as the flow field is only driven by the constant velocity $u_0$. Note that $F_z$ is negative because the hydrofoil is mounted upside down. $F_x$, respectively $F_z$, are coincident with the steady drag, respectively lift, i.e., by Eq.~\eqref{Force_eq} with $\alpha=0^\circ$, $\beta=\alpha_0=10^\circ$ and $u=u_0$. The coefficients $C_d=0.04$ and $C_l=1.17$ are obtained using Xfoil~\citep{drela1989xfoil} with a steady angle of attack $\alpha_0$. To calculate these coefficients a transition to turbulence in the boundary layer is assumed to occur near the leading edge (specifically, at 5$\%$ of the chord length) due to the roughness of the hydrofoil surface. 
Knowing the value of $C_d$ for the hydrofoil enables, by subtraction in this steady test case, an estimate the viscous friction drag $F_f=1.3$~N added by the end plates. This value agrees with the order of magnitude of the friction drag generated by rough plates of similar dimension parallel to a flow~\citep{white1991viscous}. For simplicity, we assume that the viscous drag on the end plates varies minimally under wave conditions, and take its value to be constant. The lift coefficient $C_l$ obtained using XFoil agrees well with the experimental value of 1.11 found here for $\alpha_0=10^\circ$. 

Adding waves to the steady flow generates, as expected, variations over time of the force components $F_x$ and $F_z$ that follow the wave periodicity. The larger the wave amplitude is, the greater the magnitude of the force fluctuation. Only small linear force variations appear for small incident waves (blue curves), whereas large nonlinear variations with the emergence of local peaks are observed for waves of large amplitude (red curves). During these variations, a competition occurs between drag, lift and inertia forces, which are driven by the velocity magnitude, the angle of attack and the flow acceleration. As shown in Fig.~\ref{lin_field}, the flow velocity is in phase with the wave amplitude, whereas both the flow acceleration and the angle of attack have an offset of $\pi/2$. In our case, as the hydrofoil is upside down, the maximum angle of attack is achieved on the descending slope of the waves where the deceleration of the flow is maximum.

We observe that $F_x$ has a $\pi/2$ offset with $\eta$, its minimal values are achieved when the deceleration and the angle of attack are maximum and its maximal values when the acceleration are maximum and the angle of attack minimum. This implies that the fluctuations are mostly driven by inertial effects, i.e, added-mass, consistent with the small Keulegan–Carpenter number $KC\in[0.15,3.97]$. In particular, the importance of inertia makes the horizontal force component highly negative during the flow deceleration, despite a flow moving in the positive direction. We can also note that $F_x$ does not vary symmetrically around its stationary value of $F_x\approx2.1$~N and deviates more at its minima compared to its maxima. This is due to the angle of attack that, when it is maximal, increases the value of $C_{mx}$ and so inertia effects, which here occur during the flow deceleration, and decreases $F_x$. Temporal variation of the vertical force component $F_z$, is close to being in phase opposition with $\eta$, meaning that the larger the flow velocity is, the smaller $F_z$ is. The shift observed between $F_z$ and $\eta$ is due to the angle of attack that increases lift effects on the descendant wave slope. This indicates that $F_z$ is mostly driven by the lift (which is negative) and influenced by the angle of attack. Note also that for the highest wave steepness presented, $\epsilon=0.06$, $F_z$ does not experience a single maximum each wave period but remains elevated (i.e., at its minimal lift) for a longer fraction of the wave period. The hydrofoil responses to different wave forcing are also quantified via probability density functions (PDF) in Supplemental Material, in Fig.~S3.

\begin{figure}[h!]
    \centering
    \includegraphics[width=1\linewidth]{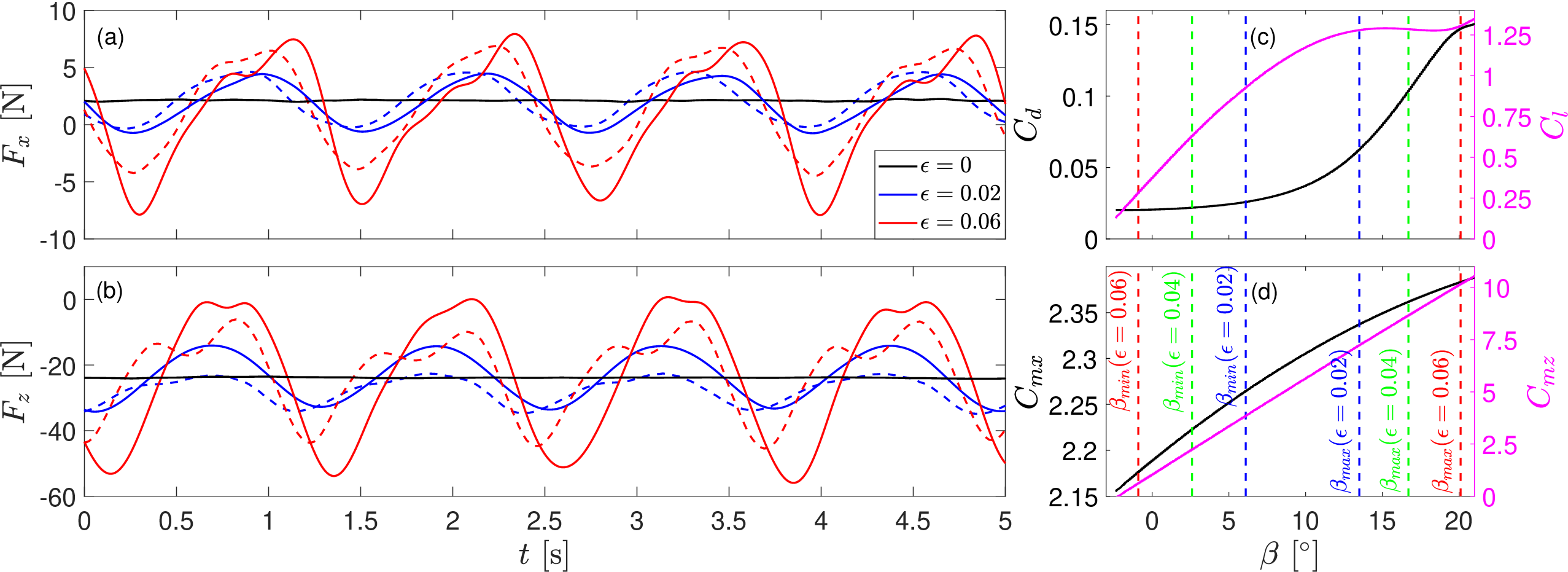}
    \caption{Theoretical model of the force variations (a) $F_x(t)$ and (b) $F_z(t)$. Experimentally measured values are plotted as solid lines and the theoretical prediction as dashed lines, for selected cases $\lambda=4$~m and $\epsilon=0$ (black), $\epsilon=0.02$ (blue) and $\epsilon=0.06$ (red). (c,d) Variation of the steady state drag and lift coefficients obtained using Xfoil, as well as the added mass coefficients obtained via potential theory, with the angle of attack $\beta$. The vertical dashed lines are indicative of the minimum and maximum angles of attack encountered in the different cases $\epsilon=0.02$ (blue), $\epsilon=0.04$ (green) and $\epsilon=0.06$ (red).}
    \label{FxFz_theory}
\end{figure}

A theoretical estimate, using Eq.~\eqref{Force_eq}, of the time-varying force components $F_x(t)$ and $F_z(t)$ is shown in Fig.~\ref{etaFxFzNotub}. 
For the purposes of this estimate, the values of the external velocity $u$ and of its orientation $\alpha$ are measured by PIV as previously described in section~\ref{setup_section}. Then, the steady time-averaged coefficients of drag and lift $C_d$ and $C_l$ that vary with $\alpha$ are obtained using Xfoil~\citep{drela1989xfoil} and plotted in Fig.~\ref{FxFz_theory}(c). The coefficients of inertia $C_{mx}$ and $C_{mz}$, that include Froude–Krylov and added mass forces~\citep{faltinsen1993sea}, are estimated using potential flow theory around a cylinder and the K\'arm\'an–Trefftz transform~\citep{milne1973theoretical} to approximates the NACA4415 closely. They are plotted as a function of the angle of attack $\beta$ in Fig.~\ref{FxFz_theory}(d). The theoretically predicted value of $C_{mx}$ at the baseline angle of attack is close to the value found experimentally during an acceleration of the hydrofoil along the water tank. The theoretical force predictions obtained from Eq.~\eqref{Force_eq} are then plotted in Fig.~\ref{FxFz_theory}(a-b) (dashed lines) and compared with the experimental data (solid lines) for weak and strong forcing cases, $\epsilon=0.02$ and 0.06. The order of magnitude of the force variation is well captured by the model, especially for the weak forcing, but is not fully in agreement with the measurements. The discrepancy stems mainly from the estimation of the coefficients. The numerical estimations of $C_d$ and $C_l$ are indeed based on steady cases, and do not take into account the dynamics present in our system. Furthermore, the estimation of the  inertia coefficients is based on linear potential flow theory, which is known to not be valid at long time and to not accurately represent the physics in stalled conditions~\citep{chaplin1997large,grift2019drag,reijtenbagh2023drag}, where nonlinear effects occur, which is expected under large amplitude wave forcing. The different components (drag + lift and inertia) from Eq.~\eqref{Force_eq} are plotted in Supplemental Material, in Fig.~S4. These evidence how inertia effects dominate the minimum value of $F_x$ and how lift (and a small drag component) drive the variation of $F_z$.  
Further numerical, experimental and theoretical works, that are beyond the scope of the present study, would be needed to properly define the coefficients and accurately model the force variations in this way.

\begin{figure}[h!]
    \centering
    \includegraphics[width=1\linewidth]{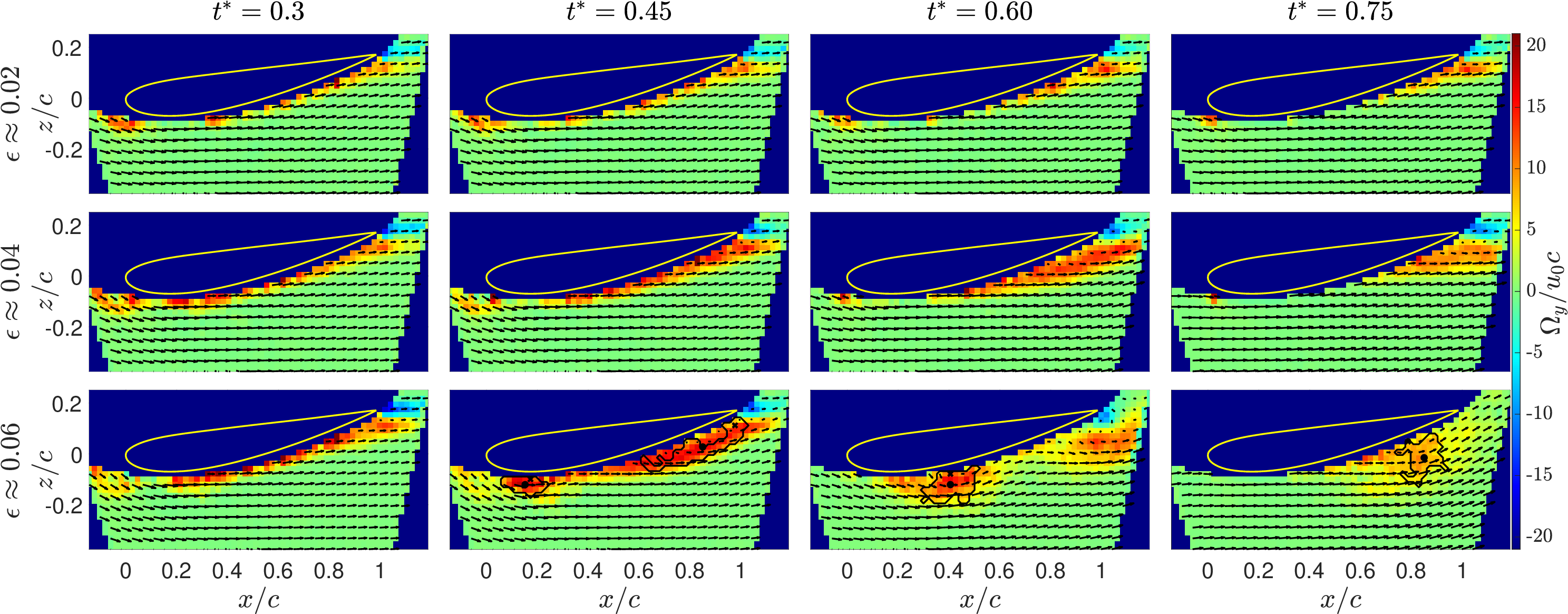}
    \caption{Phase-averaged spanwise vorticity $\Omega_y$ at selected non-dimensional times $t^*=0.3$, 0.45, 0.6, and 0.75 (from left to right), for $\lambda=4$~m and wave amplitudes $\epsilon=0.02$, 0.04, and 0.06 (from top to bottom). The hydrofoil here has a straight leading-edge and is not equipped with tubercles. Black solid lines represent isocontours of $Q$-criterion at an arbitrarily chosen value $Q=100$, which thus enclose regions of high vorticity, which we consider as vortices.  
    Black cross: centre of vortex rotation (maximum of $\Gamma_1$~\citep{michard1997identification}), black dot: geometric centroid of the vortex. Every second velocity vector is skipped, for clarity.}
    \label{PIVNotub}
\end{figure}

What is clear from the discrepancies observed in Fig.~\ref{FxFz_theory} is that linear flow theory fails to capture the intricacies of the unsteady hydrodynamics of a hydrofoil under wave forcing. This motivates an in-depth experimental characterisation of the unsteady flow which produces these forces.  
Contours of the phase-averaged spanwise vorticity, $\Omega_y$, are plotted for several times $t^*\in[0.3,0.6]$ for the case of $\lambda=4$~m and wave amplitudes $\epsilon=0.02$, 0.04, and 0.06 in Fig.~\ref{PIVNotub}. The corresponding phase-averaged velocity fields are represented by black arrows. Areas of high positive (anticlockwise) vorticity are shown in red (negative in blue) and the $Q$-criterion is used to discriminate between shear and vortices. Black solid curves represent the contours of the $Q$-criterion, set as $Q=100$ (threshold chosen to avoid noise influence), and enclose regions of positive $Q$-criterion, which we refer to as vortices. Any areas of $Q>100$ smaller than 2.3 cm$^2$ are attributed to measurement noise and not taken into account. The centre of vortex rotation (black cross) as well as the geometric centroid of the vortex (black dot) are plotted for every vortex and discussed in Section~\ref{vortex_section}. The time instants represented in this figure correspond to key moments in the development of stall. 
The full temporal evolution of the velocity and vorticity fields for this case and other forcing can be seen in the supplementary materials $movie1-8$, in which all time instants $t^*\in[0,1]$ are shown. In the interval $t^*\in[0,0.3]\cup[0.6,1]$, no relevant variation of the flow occurs as the angle of attack is close to 0$^\circ$. When $\epsilon=0.02$ (top row), no signs of boundary layer separation appear, 
as the angle of attack variations are small. 
Increasing the wave steepness to $\epsilon=0.04$ (middle row), generates angles of attack of greater amplitude, and thus flow velocity, fluctuations that induce a thickening of the vortical region near the trailing edge corresponding to the early stage of stall. Such trailing-edge boundary layer thickening and incipient separation is typical of quasi-steady stall phenomena. Finally, when the forcing amplitude is large, $\epsilon=0.06$ (last row), a compact accumulation of vorticity is observed near the leading edge. Significant boundary layer separation and roll-up is simultaneously observed near the trailing edge. The two vortices emerge at approximately $t^*=0.45$, when the angle of attack is large, and are then advected along the hydrofoil while remaining close to its surface. These observations indicate an intermediate type of stall which is not (quasi-)static in nature, but also does not correspond to a deep dynamic stall regime characterised by a single large leading-edge vortex, as is the case in~\citet{hrynuk2020effects} and \citet{valls2025dynamic}. This deep dynamic stall regime is not observed on the straight-leading-edge hydrofoil in any of the wave forcing cases studied here, because the angle variation imposed by the waves is relatively limited. 

\section{Effects of leading-edge tubercles}\label{tubercles_section}
\begin{figure}[h!]
    \centering
    \includegraphics[width=1\linewidth]{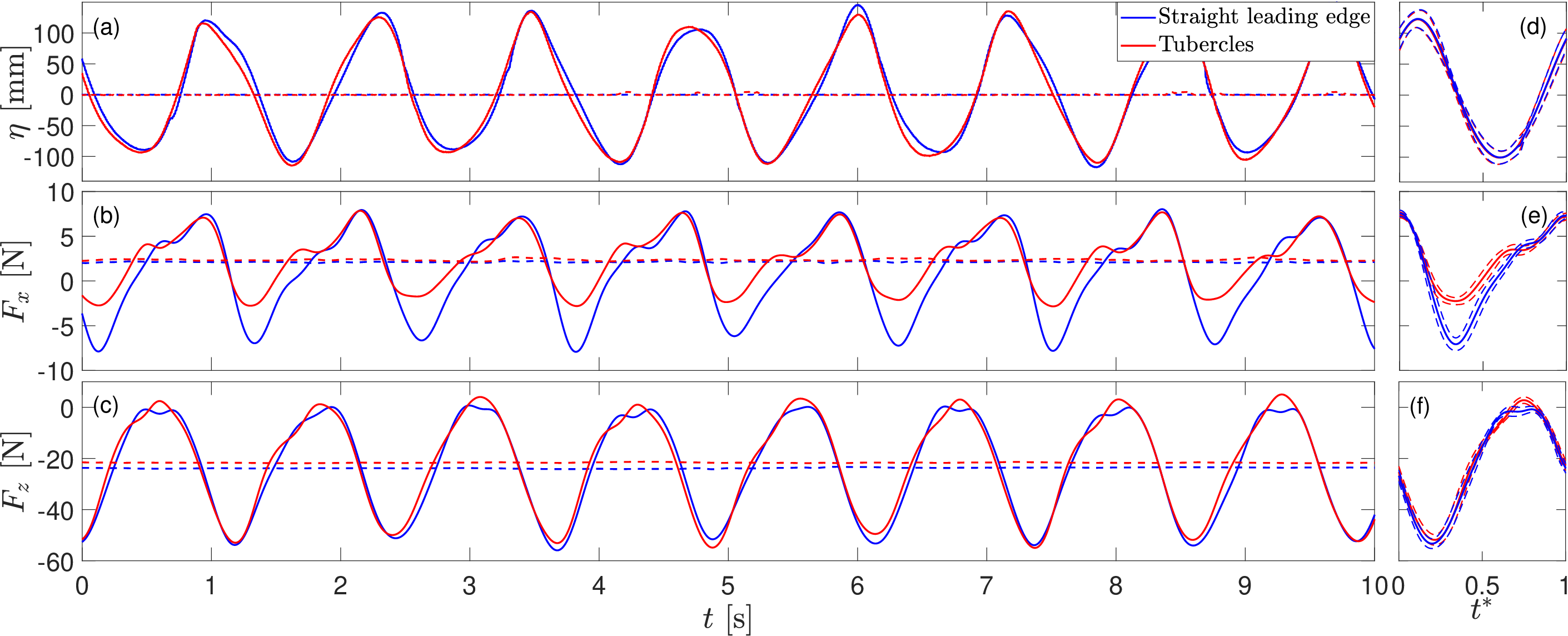}
    \caption{Time history of the surface elevation $\eta(t)$ (a), the horizontal force component $F_x$ (b), and the vertical force component $F_z$ (c), for waves of wave length $\lambda=4$~m and steepness $\epsilon=0.06$. Data are given for a hydrofoil with straight leading-edge (blue solid lines) and a hydrofoil with leading-edge tubercles (red solid lines). (d,e,f) Phase-averaged data over 40 wave periods of the same. Dashed lines correspond to the standard deviation of the average.}
    \label{etaFxFzNotubTub}
\end{figure}

In this section, the hydrodynamic behaviour of our second model, i.e., a hydrofoil equipped with leading edge tubercles, is tested and compared with the straight-leading-edge case. The tubercled geometry is expected to modify the flow behaviour, including the dynamic stall characteristics~\citep{hrynuk2020effects,valls2025dynamic}. The effect of tubercles on the generation of hydrodynamic force is quantified in 
Fig.~\ref{etaFxFzNotubTub}, where the surface elevation $\eta(t)$ and the horizontal and vertical force components, $F_x(t)$ and $F_z(t)$, are plotted over time for each of the two geometries. The steady case with no waves (dashed lines) and a selected wave-forced case, with $\lambda=4$~m and $\epsilon=0.06$ (solid lines), are shown. 
The same quantities, phase averaged over 40 wave periods (solid lines) along with the corresponding standard deviation (dashed lines), are plotted in Fig.~\ref{etaFxFzNotubTub}(d-f). This large amplitude case is plotted here because it corresponds to conditions where the effects of tubercles are most visible. Other weaker wave forcing cases are plotted in the Supplemental Material in Fig.~S5 and S6. For the steady case, with no waves, $F_x\approx 2.1\pm0.1$~N for the straight-leading-edge case and $F_x^{tub}\approx 2.4\pm0.1$~N for the tubercled hydrofoil, meaning that tubercles increase the steady-state drag by about 14$\%$, which is relatively small compared with the magnitude of the $F_x$ fluctuations under wave forcing. The modification of the geometry therefore does not greatly impact the horizontal component of the force. Regarding the vertical force component, $F_z\approx -23.8\pm0.2$~N and $F_z^{tub}\approx -22.1\pm0.2$~N, so an increase (i.e.\ decreased lift) of about $7\%$ is observed with tubercles. These two observations, i.e., small increase of drag and small decrease of lift, correspond to what has been observed before in steady cases for angles of attack smaller than the stall angle~\citep{johari2007effects,fan2022numerical}.

\begin{figure}[h!]
    \centering
    \includegraphics[width=1\linewidth]{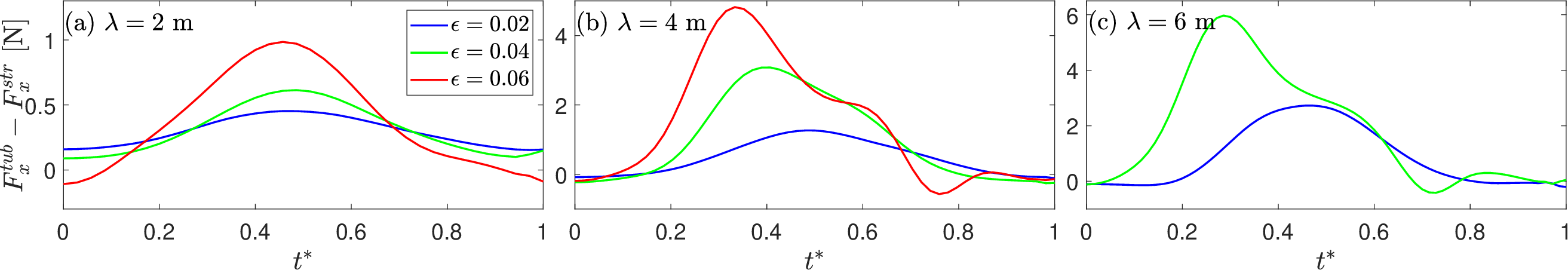}
    \caption{Variation over the non-dimensional time $t^*$ of the difference between the horizontal hydrodynamic force component in the cases with $F_x^{tub}$ and without $F_x^{str}$ tubercles, for (a) $\lambda=2$~m, (b) $\lambda=4$~m and (c) $\lambda=6$~m, for wave steepness $\epsilon=0.02$ (blue), $\epsilon=0.04$ (green) and $\epsilon=0.06$ (red). Note the differing scales of the vertical axes.}
    \label{FxTub-FxNoTub_t}
\end{figure}

Under unsteady wave forcing, just as in the straight-leading-edge case, variation of the force components $F_x$ and $F_z$ emerges, following the wave periodicity. A comparison with the hydrodynamic forces generated in the straight-leading-edge case reveals that the primary effect of tubercles appears in the $F_x(t)$ component [Fig.~\ref{etaFxFzNotubTub}(b,e)]. Fluctuations of $F_x(t)$ are attenuated by approximately $35\%$ when leading-edge tubercles are introduced. This effect is quantified for several wave conditions in Fig.~\ref{FxTub-FxNoTub_t}, where the difference $F_x^{str}-F_x^{tub}$ between the horizontal forces for the two hydrofoils is plotted over the non-dimensional time $t^*$. The force difference becomes non negligible (deviation above $15\%$ of the initial value) when $t^*\in[0.2,0.8]$, which corresponds to the period during which the hydrofoil experiences a large angle of attack and where stall occurs. Outside of this interval, where the angle of attack is small, the flow remains attached to the hydrofoil and $F_x^{str}-F_x^{tub}\approx0$~N. This implies that the hydrodynamic effect of tubercles under wave-induced forcing is primarily driven by their mediation of hydrofoil stall dynamics. 
It is of practical interest to note, that the observed reduction in force fluctuation magnitude could be beneficial with regard to reducing structural fatigue, but at the cost of an increased mean drag. 
The impact of tubercles on horizontal forces generated during dynamic stall has, to our knowledge, never been previously investigated, as former studies focused primarily on vertical forces~\citep{hrynuk2020effects,valls2025dynamic}.

We observe only a minor impact of the presence of tubercles on the vertical force component, $F_z$. The strong increase in lift expected at large angle of attack in steady cases~\citep{johari2007effects,fan2022numerical} is not observed here. The effect of tubercles on lift generation under dynamic stall has not previously been measured. In~\citet{hrynuk2020effects} forces were not directly measured, and the authors only assumed a tubercle-induced increase of lift based on a greater measured circulation in the leading-edge stall vortex. Although our measurements agree with~\citet{hrynuk2020effects} that circulation in the leading-edge vortex is increased by the presence of tubercles, we show that this does not translate to an increased lift force. 
Indeed, in~\citet{valls2025dynamic}, numerical simulations indicate that tubercles decrease the peak lift experienced by a foil undergoing dynamic stall induced by a linear pitch-up manoeuvre. This, they attribute to a decrease in circulation which was experimentally observed in the same paper. The apparent inconsistency between the two papers might be due to the differing pitch rate ranges they investigated ($\Omega^*>0.1$ in \citet{hrynuk2020effects} and $\Omega^*=0.05$ in \citet{valls2025dynamic}), although the absence of direct force measurements in either study prevents definitive conclusions. In the present study, the effect of tubercles on the vertical force is minimal, with the exception of a small deviation near the maximal values of $F_z$: The vertical force in the straight-leading-edge case peaks twice per wave period, while with tubercles the first peak is attenuated and the second enhanced, resulting in a single, asymmetric peak in $F_z$. Because of the slight period-to-period variability in the wave forcing, this deviation in $F_z$ is best observed via the phase-averaged force curves in Fig.~\ref{etaFxFzNotubTub}(d-f) as well as in probability density functions (PDF) of the measured force signals, which are included in Supplemental Material, in Fig.~S7.

\begin{figure}[h!]
    \centering
    \includegraphics[width=1\linewidth]{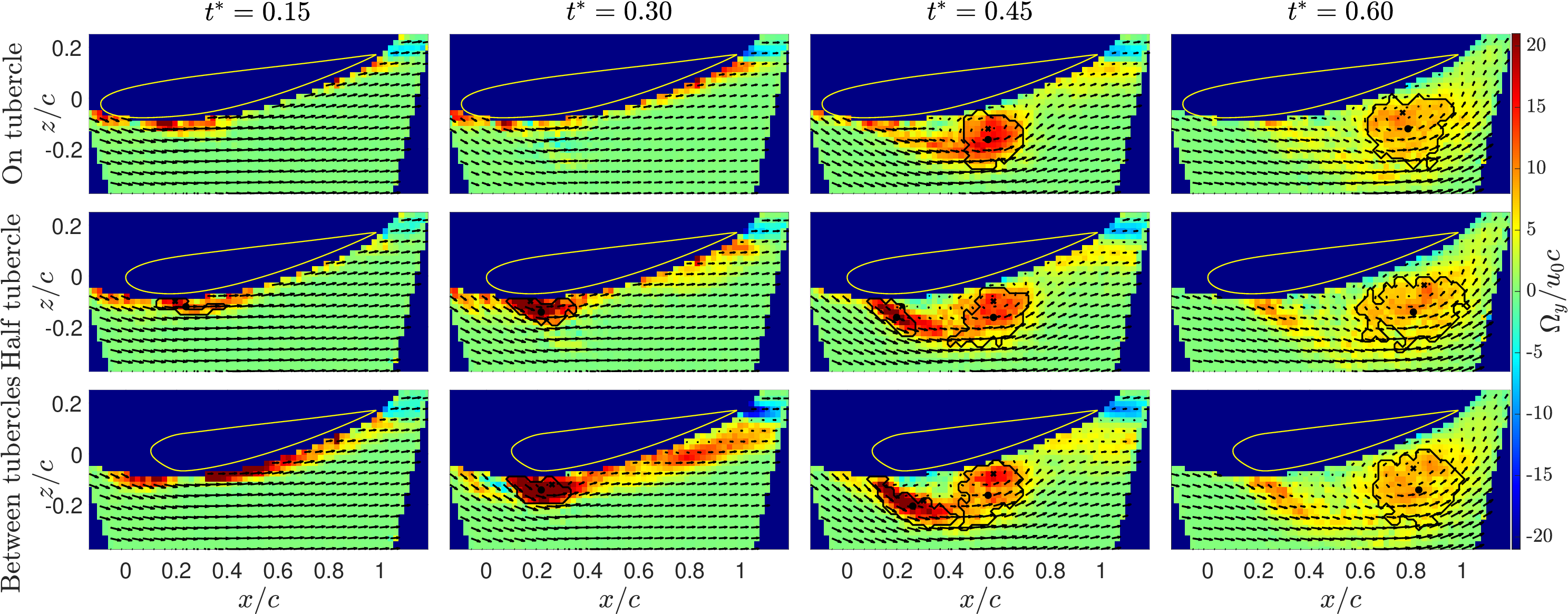}
    \caption{Phase-averaged spanwise vorticity $\Omega_y$ below a hydrofoil fitted with leading-edge tubercles, at selected non-dimensional times $t^*=0.15$, 0.3, 0.45, and 0.6 (from left to right), for $\lambda=4$~m and wave amplitude $\epsilon=0.06$. Three separate measurement planes are shown: at a tubercle peak, halfway between tubercle peak and trough, and at the trough between two tubercles (from top to bottom). Black solid lines represent isocontours of $Q$-criterion at an arbitrarily chosen value $Q=100$, which thus enclose regions of high vorticity, which we consider as vortices. Black cross: centre of vortex rotation (maximum of $\Gamma_1$~\citep{michard1997identification}), black dot: geometric centroid of the vortex. Every second velocity vector is skipped, for clarity.}
    \label{PIVTub}
\end{figure}

To better understand how the flow is affected by tubercles, the phase-averaged spanwise vorticity $\Omega_y$ around the hydrofoil for the case of wave length $\lambda=4$~m and large wave steepness $\epsilon=0.06$ is plotted in Fig.~\ref{PIVTub} for selected dimensionless times $t^*\in[0.15,0.6]$, corresponding to the instants of interest. The three planes in which measurements were taken (at a tubercle peak, halfway between tubercle peak and trough, and at the trough between two tubercles) are represented in the figure (from top to bottom). Comparing the vorticity contours in this figure to those in Fig.~\ref{PIVNotub} demonstrates that the flow is strongly altered by the presence of tubercles. The corresponding videos over the complete wave period $t^*\in[0,1]$, including for cases of weaker wave forcing amplitude and larger wave length, can be found in $movie1-8$. A concentrated region of spanwise vorticity first emerges near the leading edge in the planes of measurement located between the tubercles and halfway between tubercle peak and trough (Fig.~\ref{PIVTub}, bottom two rows). At $t^*=0.3$, the leading-edge vortex has not yet appeared in the tubercle peak measurement plane, indicating that the initial boundary layer separation and stall vortex roll-up are localized phenomena associated with the tubercle troughs and the stall vortex is, in its early stages, spanwise discontinuous. 
By $t^*=0.45$ the leading-edge vortex appears in all measurement planes, and remains at this stage attached to the leading edge via a strong shear layer. 
This large vortex continues to grow over time, both spatially and in strength, and then detaches and is advected along the hydrofoil from approximately $t^*=0.6$. The leading-edge vortex generated in the tubercled case is much greater in spatial extent than that observed in the straight-leading-edge case.
Because of the limited angle-of-attack variation amplitude ($20^\circ$ for $\lambda=4$~m and $\epsilon=0.06$ in our most extreme forcing case, against $30^\circ$ in \citet{valls2025dynamic} and $50^\circ$ in \citet{hrynuk2020effects}) in the present work, deep stall is not fully achieved in the case of straight-leading-edge geometry, placing our experiments close to a quasi-static stall regime. When tubercles are applied, however, behaviour similar with the other studies occurs, i.e., vortex generation initiates between tubercles at smaller angle of attack and leads to the generation of a more diffuse large leading-edge vortex structure. This means that, although the angle-of-attack variation is here too small to trigger a deep stall regime on the straight-leading-edge hydrofoil, the presence of tubercles initiates early dynamic stall at the leading edge, leading eventually to a large leading-edge vortex present across the span. 

\section{Analysis of the vortex dynamics}\label{vortex_section}
\begin{figure}[h!]
    \centering
    \includegraphics[width=1\linewidth]{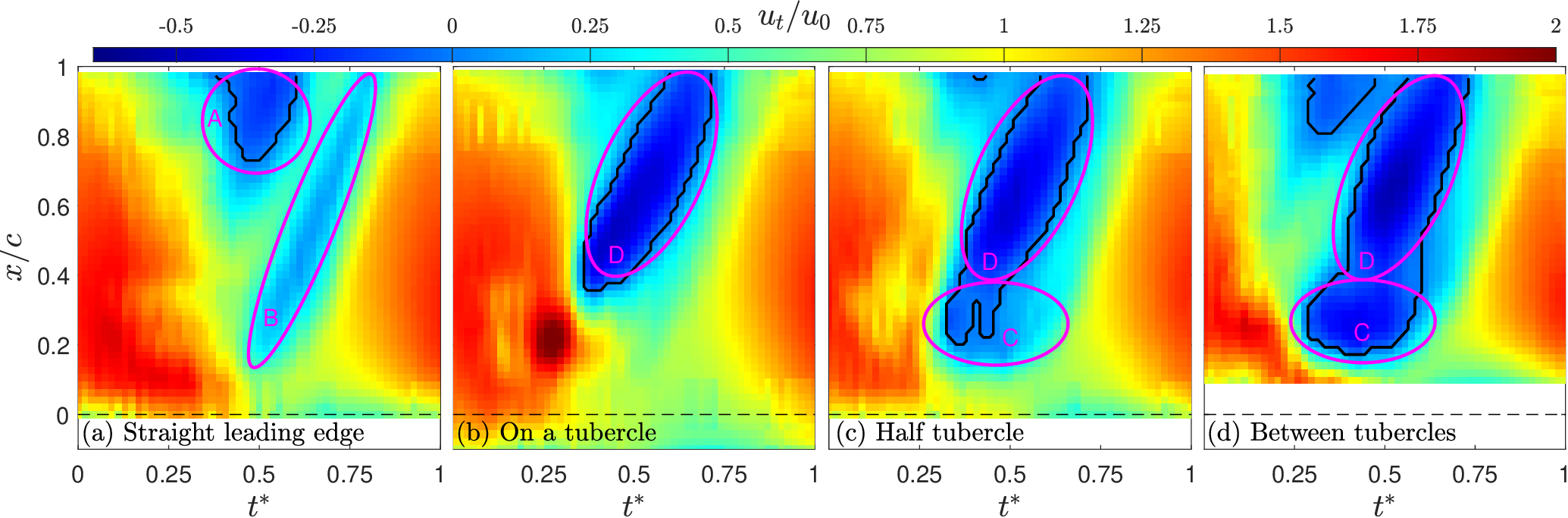}
    \caption{Spatio-temporal variations of the non-dimensional phase-averaged velocity, $u_t/u_0$, tangent to the hydrofoil suction surface at non-dimensional times $t^*\in[0,1]$ for $\lambda=4$~m and $\epsilon=0.06$. The two cases (a) straight leading edge and (b-d) with tubercles are shown. (b) shows the measurement plane on a tubercle peak, (c) halfway between tubercle peak and trough, and (d) at the trough between tubercles. Black solid curves delimit spatio-temporal regions in which $u_t<0$. Alphabetically-labelled magenta ellipses indicate zones of interest to ease description.}
    \label{Utan_X}
\end{figure}

This section presents a more quantitative in-depth analysis of dynamic stall vortex initiation and evolution on a hydrofoil under action of waves, and examines the effect of leading-edge tubercles on the dynamic stall process. We first examine the tangential velocity of the flow along the suction surface of each hydrofoil, defined as $u_t=\textbf{u}\cdot\textbf{t}$ where $\textbf{t}$ are the vectors locally tangent to the hydrofoil suction surface. 
The values of $u_t$ are plotted as a function of space and time in Fig.~\ref{Utan_X} for (a) the straight-leading-edge hydrofoil, and (b-d) the tubercled hydrofoil, for the case of $\lambda=4$~m and $\epsilon=0.06$. Black solid contours delimit the spatio-temporal locations at which reverse flow occurs, i.e., where $u_t<0$. Discrete instances of reverse, or near-reverse, flow are marked using magenta ellipses, and labelled alphabetically to ease description. Other wave forcing, $\lambda=4$~m, $\epsilon=0.04$ and $\lambda=6$~m, $\epsilon=0.04$ are shown in Supplemental Material, in Fig.~S8 and S9, and display similar results.

During the time intervals $t^*\in[0,0.2]\cup[0.8,1]$, which correspond to angles of attack close to $0^\circ$, $u_t$ is similarly distributed (correlation coefficient approximately 0.8) for the two hydrofoils: A streamwise acceleration from the leading edge to a maximum value of $u_t\approx 1.25$ m/s at approximately $x/c\approx 0.25$, and then subsequent deceleration. No reverse flow is present during these time intervals. In the straight-leading-edge case, a reverse flow first appears at the trailing edge in the zone marked ``A''. This is followed by the emergence of a nearly reversed flow (zone B) near the leading edge, which subsequently propagates toward the hydrofoil's trailing edge. It should be noted that, because of the finite resolution of the PIV measurement, $u_t$ is obtained at approximately 4 mm offset from the actual surface of the hydrofoil, and the boundary layer is not well resolved. As such, values of $u_t$ are indicative of the approximate magnitude of the outer flow velocity, and its direction, but may not accurately capture very localised, small-scale, flow reversal within the boundary layer. 
This means that there is uncertainty as to whether flow reversal actually occurs or not in zone B. The initial onset of flow reversal near the trailing edge is typical of static, or quasi-static, stall phenomena~\citep{buchner2018stall}, which is not unexpected due to the low dimensionless pitch rates $\Omega^*\in[0.01,0.06]$ studied here. The subsequent leading edge (near-)reversal and associated vortex roll-up (Fig.~\ref{PIVNotub}) indicate, however, that incipient dynamic stall behaviour is also present.

As seen in Section~\ref{tubercles_section}, the addition of tubercles modifies the flow. First, for a short instant, $u_t$ is observed to increase on tubercles at $t^*=0.25$ and $x/c=0.2$ [Fig.~\ref{Utan_X} (b)]. At the same streamwise location, a reverse flow emerges very shortly thereafter between the tubercles [zone C in Fig.~\ref{Utan_X} (d)]. 
The emergence of this reverse flow is presumably triggered by three-dimensional flows that correspond to streamwise vortices~\citep{valls2025dynamic}. 
The reverse flow of zone C corresponds to the accumulation of vorticity observed previously in Fig.~\ref{PIVTub} and remains at an approximately constant $x/c$ location until $t^* \approx 0.6$. This relatively stationary leading-edge-attached vorticity accumulation, and associated reverse flow region, is distinct from the primary leading-edge vortex structure, and appears related to the shear layer connecting the leading-edge vortex to the leading edge. The signature of the leading-edge vortex on the surface-tangential velocity is observed in a distinct region (zone D) of reverse flow,  
which presents at all measured planes along the spanwise direction 
from $t^*\in[0.4,0.75]$. In this time interval, the reverse flow in zone D is observed to advect downstream at an approximately constant rate. The observed behaviour of zone D is qualitatively similar to the slightly earlier evolution of zone B in the straight-leading-edge case, but of a clearly larger magnitude. 
Note also the presence of a trailing-edge separation observed in Fig.~\ref{Utan_X} (d), which is qualitatively similar to the quasi-static stall observed in the straight-leading-edge case [zone A in Fig.~\ref{Utan_X} (a)]. This separation does not appear in the on-tubercle measurement plane [Fig.~\ref{Utan_X} (b)], as the streamwise vortices created by the tubercles induce a wall-normal flux of momentum towards the surface in that plane, but not in the plane situated between tubercles~\citep{valls2025dynamic}.

\begin{figure}[h!]
    \centering
    \includegraphics[width=1\linewidth]{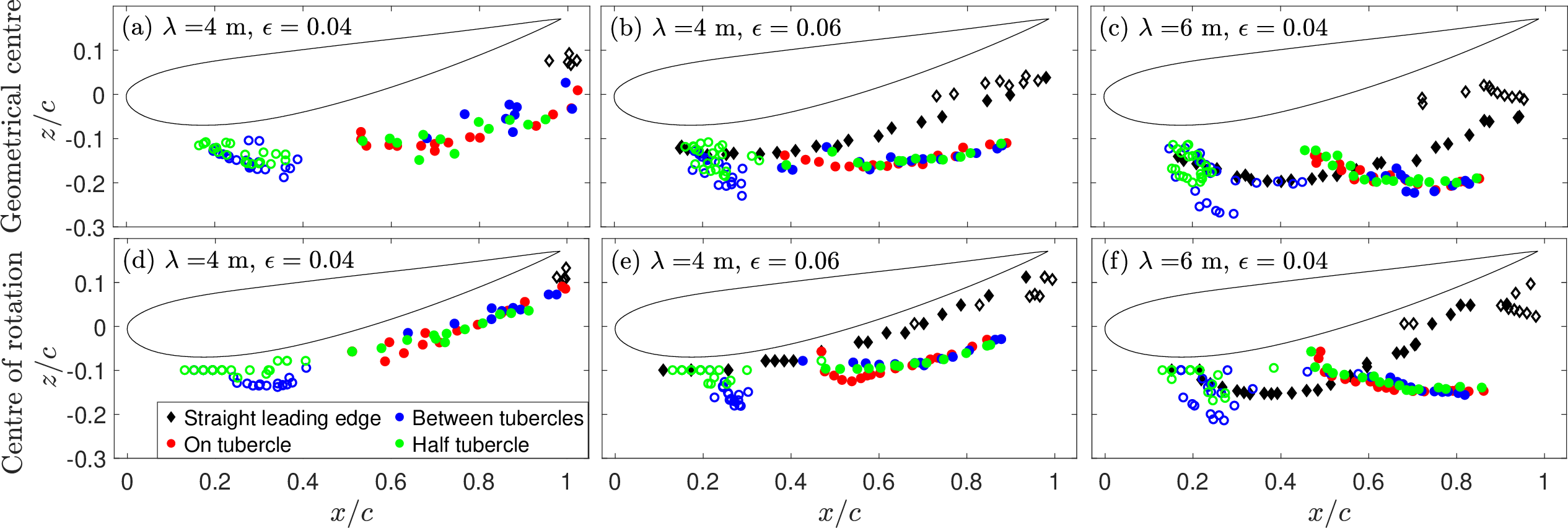}
    \caption{Location of the geometric centres (a-c) and centres of rotation (d-f) of each identified vortex, for $t^*\in[0,1]$. Three cases of wave forcing are shown: (a,d) $\lambda=4$~m, $\epsilon=0.04$, (b,e) $\lambda=4$~m, $\epsilon=0.06$, and (c,f) $\lambda=6$~m, $\epsilon=0.04$. Symbol shape and colour indicate different planes of measurement, i.e., straight leading edge (black diamonds), measurement plane coincident with a tubercle peak (red dots), measurement plane coincident with a tubercle trough (blue dots), and measurement plane halfway between tubercle peak and trough (green dots). Empty symbols correspond to the structures from zone A and C and full symbols to zones B and D. Hydrofoil shape and location are indicated via black solid outline.}
    \label{XZ_vortex}
\end{figure}

Individual phase-averaged vortices are identified and tracked over time from the PIV measurement data, using a method similar to the one used in~\citet{fiscaletti2018spatial} with two separate definitions: Vortex locations can be extracted as the geometric centres of regions in which $Q>100$ (black dots in Fig.~\ref{PIVNotub} and \ref{PIVTub}), or the locations can alternatively be found using the $\Gamma_1$ method, described by \citet{michard1997identification}, which computes the centres of rotation of the region in which $Q>100$ (black crosses in Fig.~\ref{PIVNotub} and \ref{PIVTub}). We focus here on tracking and characterizing those vortex structures with an origin at leading edge or in the suction-surface boundary layer: The clockwise-rotating trailing edge vortex that appears in some large-amplitude test cases is not considered in this analysis. 
The tracked vortex locations, defined by the geometric centres, are plotted in Fig.~\ref{XZ_vortex}(a-c), and as defined by the centres of rotation in Fig.~\ref{XZ_vortex}(d-f), for three separate wave forcing cases: (a,d) $\lambda=4$~m, $\epsilon=0.04$, (b,d) $\lambda=4$~m, $\epsilon=0.06$, and (c,f) $\lambda=6$~m, $\epsilon=0.04$. These three cases correspond to wave forcing of sufficient magnitude such that stall vortices can form and be detected in our measurement. Each marker's shape and colour indicates the leading-edge geometry and measurement plane to which it relates. Empty symbols relate to the vortex structures associated with zones A and C, and full symbols to the vortex structures associated with zones B and D that are defined in Fig.~\ref{Utan_X}. In the case of a straight leading edge (black diamonds), almost no vortex emerges for $\lambda=4$~m, $\epsilon=0.04$ (a,d), as the maximum achieved angle of attack of $17^\circ$ in this case is too small to generate measurable dynamic stall over the time scale of wave forcing, only the boundary layer thickening and roll-up near the trailing edge (zone A, empty diamonds) tends to emerge. Increasing the wave steepness to $\epsilon=0.06$ (b,e) produces the quasi-static stall case where flow separation, boundary layer roll-up, and resulting vortex structures occur near both the trailing edge (zone A, empty diamonds in $x/c\in[0.8,1]$) and the leading edge (zone B, full diamonds in $x/c\in[0.2,1]$). 
Wave forcing at longer wave length, $\lambda=6$~m (c,f), implies larger variation of the angle of attack leading to stronger reverse flow and therefore stall. When tubercles are present at the leading edge, vorticity accumulation near the leading edge at $x/c\approx0.2$ (zone C) is detected, in all plotted cases, in between tubercles (blue empty markers) and in the measurement plane halfway between tubercle trough and peak (green empty markers). The emergence of the large leading-edge vortex, indicating a dynamic stall regime, is detected in all measurement planes (full blue, green and red markers) at $x/c\in[0.4,1]$. This vortex forms farther from the hydrofoil leading edge, at about $x/c\approx0.45$, than in the straight-leading-edge hydrofoil wherein it forms at about $x/c\approx0.2$. 
The two methods of vortex detection employed here yield similar results to one another, with the only difference being that the vortices' centres of rotation are located closer to the hydrofoil surface than their geometric centres. This indicates a vortex asymmetry, which is also clear in Fig.~\ref{PIVTub}. We select only the geometric centroid for use in subsequent analysis. 

\begin{figure}[h!]
    \centering
    \includegraphics[width=1\linewidth]{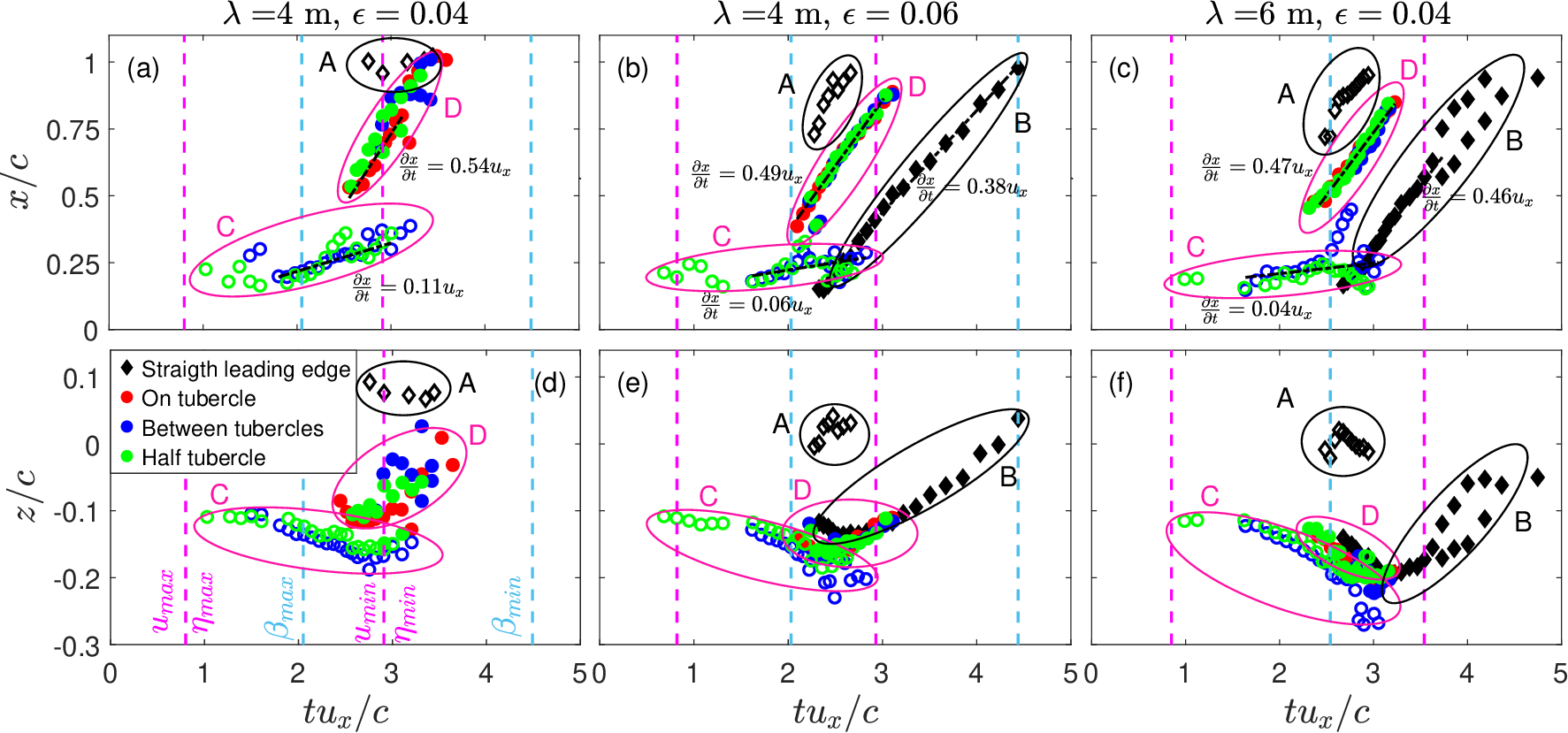}
    \caption{Variation of the $x$ (a-c) [respectively $z$ (d-f)] location of vortex geometric centroid with the non-dimensional time $tu_x/c$, for the three wave forcing cases (a,d) $\lambda=4$~m, $\epsilon=0.04$, (b,e) $\lambda=4$~m, $\epsilon=0.06$, and (c,f) $\lambda=6$~m, $\epsilon=0.04$. Symbol shape and colour indicate different planes of measurement, i.e., straight leading edge (black diamonds), measurement plane coincident with a tubercle peak (red dots), measurement plane coincident with a tubercle trough (blue dots), and measurement plane halfway between tubercle peak and trough (green dots). Black dash-dotted lines: linear best fits. Alphabetically-labelled ellipses indicate zones of interest to ease description, with labelling identical to that used in Fig.~\ref{Utan_X}. Black ellipses relate to the case of the straight-leading-edge hydrofoil, while data for tubercled hydrofoils are delineated using magenta ellipses. Empty symbols correspond to the vortices from zone A and C and full symbols to zones B and D. Vertical dashed magenta lines refer to the instants of maximal and minimal horizontal velocity $u_x$ (wave peak and trough) and the blue vertical dashed lines to the instants of maximal and minimal angle of attack $\beta$.}
    \label{XZ_vortez_t}
\end{figure}

Fig.~\ref{XZ_vortex} indicates detected vortex locations in space, for various measurement times, but without explicitly including temporal information. In Fig.~\ref{XZ_vortez_t}, the horizontal $x$ (a-c) and vertical $z$ (d-f) locations of the vortices are plotted over time, resulting in similar dynamics for the three different wave forcing cases that are shown here. Note that here the time axis is non-dimensionalised using the convective time, $c/u_x$, with $u_x$ the horizontal flow velocity measured by PIV, instead of the wave period, $T=1/f_r$. This non-dimensionalisation is appropriate here as motion of the vortices post-detachment is driven by the external flow velocity. It should be noted, however, that the use of the non-dimensionalisation $t^* = t/T$ yields similar results, as the horizontal velocity $u_x$ depends on the wave period as $u_x\sim u_0\pm\mathcal{R}\omega$. The horizontal advective velocity of the vortices is inferred by a linear fit to the experimentally observed variation of $x$ (dash-dotted lines). In the same way as in Fig.~\ref{Utan_X}, various features of interest in Fig.~\ref{XZ_vortez_t} are indicated by black (straight leading edge) and magenta (tubercled leading edge) ellipses, keeping consistent alphabetic notation. 
As was observed previously in the case of a straight leading edge, except for in the case $\lambda=4$~m and $\epsilon=0.04$, vortices are formed both near the trailing edge (zone A) and the leading edge (zone B), and are advected following roughly the hydrofoil surface. With tubercles present, the vorticity accumulation (zone C) near the leading edge appears early and remains closely associated with the leading edge for the duration of the measurement: Its motion is characterised only by a small horizontal velocity of magnitude between $0.04u_x$ and $0.11u_x$ and a slight decrease over time of its $z$ coordinate related to its spatial growth. Some time later, the large leading-edge vortex (zone D) forms and is observed to travel horizontally at a velocity of approximately $0.47\pm0.09 u_x$, similar to the advective velocity in straight-leading-edge case (zone B) despite a much earlier emergence. Except for in the case of $\lambda=4$~m and $\epsilon=0.04$, this large leading-edge vortex moves only negligibly in the vertical direction, remaining located at $z/c\approx0.1$, and does not follow the hydrofoil shape, which is contrary to the behaviour observed in the straight-leading-edge case. The difference is also easily observed in Fig.~\ref{XZ_vortex}. This observation differs from \citet{hrynuk2020effects} where tubercles make the vortex travel closer to the hydrofoil surface compared to the straight-leading-edge case. This difference could be due to the difference between using wave-induced flow motion rather than a pitching movement to trigger dynamic stall. Note also that from $tu_x/c>3.5$, the leading-edge vortex (zone D) has been advected beyond the tubercled hydrofoil's trailing edge, whereas it is still present in the straight-leading-edge case (zone B). This difference in the length of time over which the leading-edge vortex resides in the vicinity of the hydrofoil suction surface may relate to the small qualitative difference in the vertical force $F_z$ time history 
creating the two peaks observed previously in Fig.~\ref{etaFxFzNotubTub}(c) at each period. 

\begin{figure}[h!]
    \centering
    \includegraphics[width=1\linewidth]{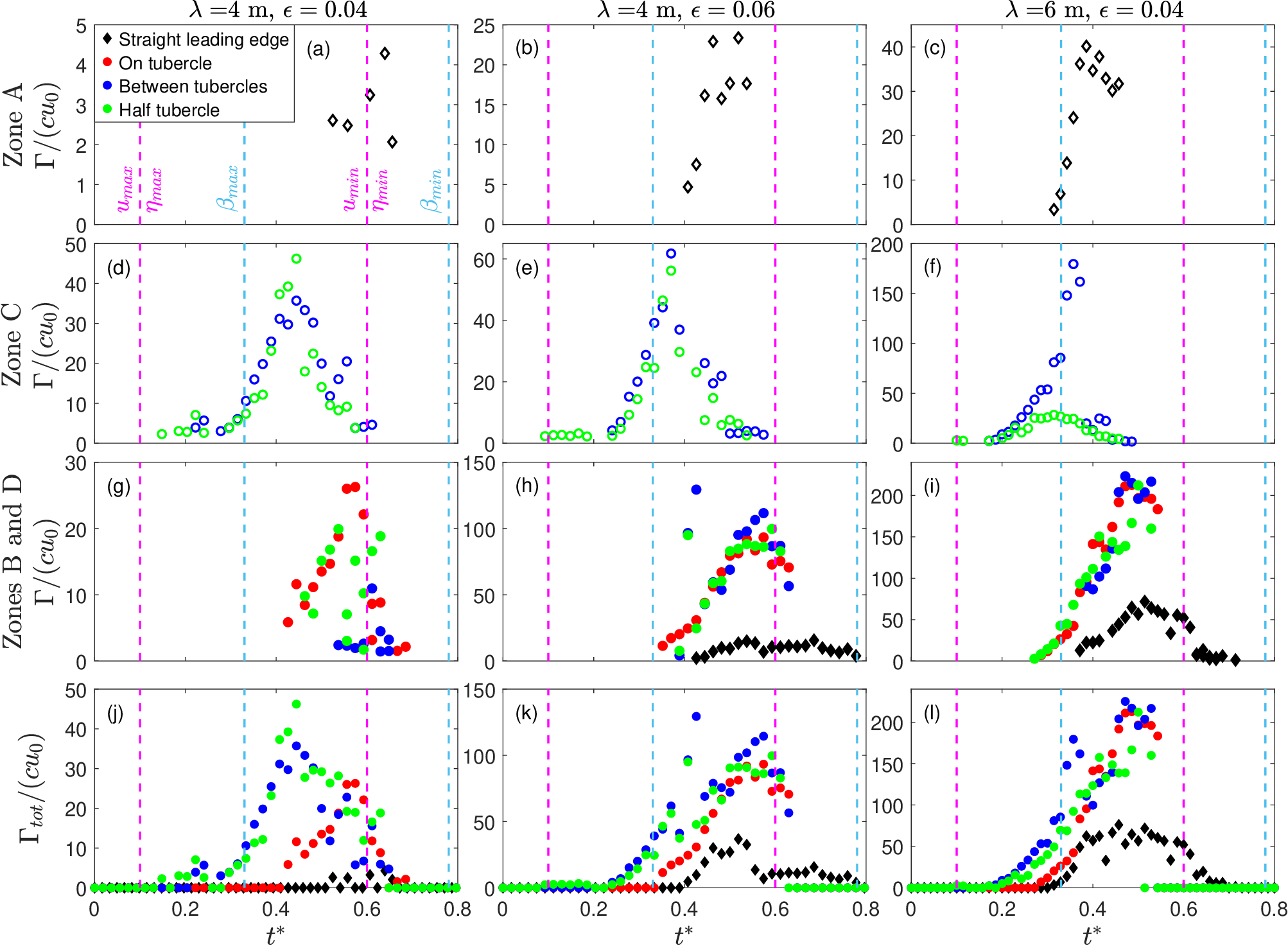}
    \caption{Evolution of the circulation $\Gamma$ as function of non-dimensional time $t^*$ for the three wave forcing cases (a,d,g,j) $\lambda=4$~m, $\epsilon=0.04$, (b,e,h,k) $\lambda=4$~m, $\epsilon=0.06$, and (c,f,i,l) $\lambda=6$~m, $\epsilon=0.04$. The different contributions are represented separately: (a,b,c) zone A, (d,e,f) zone C and (g,h,i) zones B and D and (j,k,l) the total circulation $\Gamma_{tot}$ including all contributions. Symbol shape and colour indicate different planes of measurement, i.e., straight leading edge (black diamonds), measurement plane coincident with a tubercle peak (red dots), measurement plane coincident with a tubercle trough (blue dots), and measurement plane halfway between tubercle peak and trough (green dots). Vertical dashed magenta lines refer to the instants of maximal and minimal horizontal velocity $u_x$ (wave peak and trough) and the blue vertical dashed lines to the instants of maximal and minimal angle of attack $\beta$. Note the varying y-axis scale.}
    \label{Circulation}
\end{figure}

The circulation contained within the identified vortex structures is calculated as
\begin{equation}
    \Gamma=\int_{Q>100} \Omega_y dxdz
    \label{circu_eq}
\end{equation}
\noindent and plotted for the three different wave forcing conditions over time in Fig.~\ref{Circulation}. The contributions from the different vortex structures are plotted separately, (a,b,c) zone A, (d,e,f) zone C and (g,h,i) zone B and D, whereas the total circulation $\Gamma_{tot}$ with every contribution included is reported in (j,k,l). Vortices that start to leave the measurement domain are no longer taken into account. The time-evolution of shed circulation is qualitatively similar across the various cases; only the magnitude changes, increasing with wave forcing intensity (wave length and steepness). In the straight-leading-edge case (black diamonds), circulation increases first because of the near-trailing-edge boundary layer roll-up (zone A), then subsequently because of the formation of the leading-edge vortex (zone B). The two structures, once they are fully formed, have a similar circulation magnitude to one another, and so both contribute substantially to the total circulation $\Gamma_{tot}$. In the case of the tubercled hydrofoil, circulation appears first at the measurement planes between tubercles and halfway between tubercle peaks and troughs because of the leading-edge vorticity accumulation (zone C). The circulation contained within this structure then decreases from about $t^*\approx0.35$ as the angle of attack decreases and circulation is transferred to the large leading-edge vortex 
(zone D). The rate of increase of circulation in the leading-edge vortex is similar in all three planes of measurement. This leading-edge vortex contains a much greater circulation (approximately four times greater magnitude) compared with that observed in the straight-leading-edge case (zone B) [Fig.~\ref{Circulation}(h,i)]. 
The increase in total circulation in the plane coincident with the tubercle peaks lags behind that in the other measurement planes, as the circulation in the leading-edge vortex primarily originates from flow separation in the tubercle troughs, where a stronger shear layer is observed. 
Upon reduction in the strength of the feeding shear layer in between tubercles and in the half-tubercle plane from $t^*\approx0.5$, the total circulation measured in those planes reduces gradually towards parity with the circulation in the tubercle-coincident plane. $\Gamma_{tot}$ is thereafter approximately identical for the three planes, as it consists almost entirely of circulation in the large diffuse leading-edge vortex. 
Direct comparison with previous studies on the effect of tubercles on dynamic stall~\citep{hrynuk2020effects,valls2025dynamic} is difficult due to differing parameter ranges. We nevertheless observe concurrence with \citet{hrynuk2020effects}, in that the circulation is of greater magnitude between tubercles than on a tubercle. They observe that, as time progresses, the circulation in the tubercle peak plane becomes greater, a discrepancy with our observations which might be attributed to the non-dimensional pitching rate $\Omega^*$ being more than twice in that work than in the present study. As is common in the study of steady flow around a foil, it has previously been assumed that circulation in the separated vortex structures under dynamic stall is directly linked to lift force generation
~\citep{hrynuk2020effects,valls2025dynamic}. 
In the present work, the circulation in the leading-edge vortex structures is greater in the presence of tubercles, but the time history of the vertical force, $F_z$, is similar for both leading-edge geometries [Fig.~\ref{etaFxFzNotubTub}(c)]. This contradicts earlier works' assumptions that tubercles' effects on the leading-edge stall structures would significantly impact the lift force generation.

\begin{figure}[h!]
    \centering
    \includegraphics[width=1\linewidth]{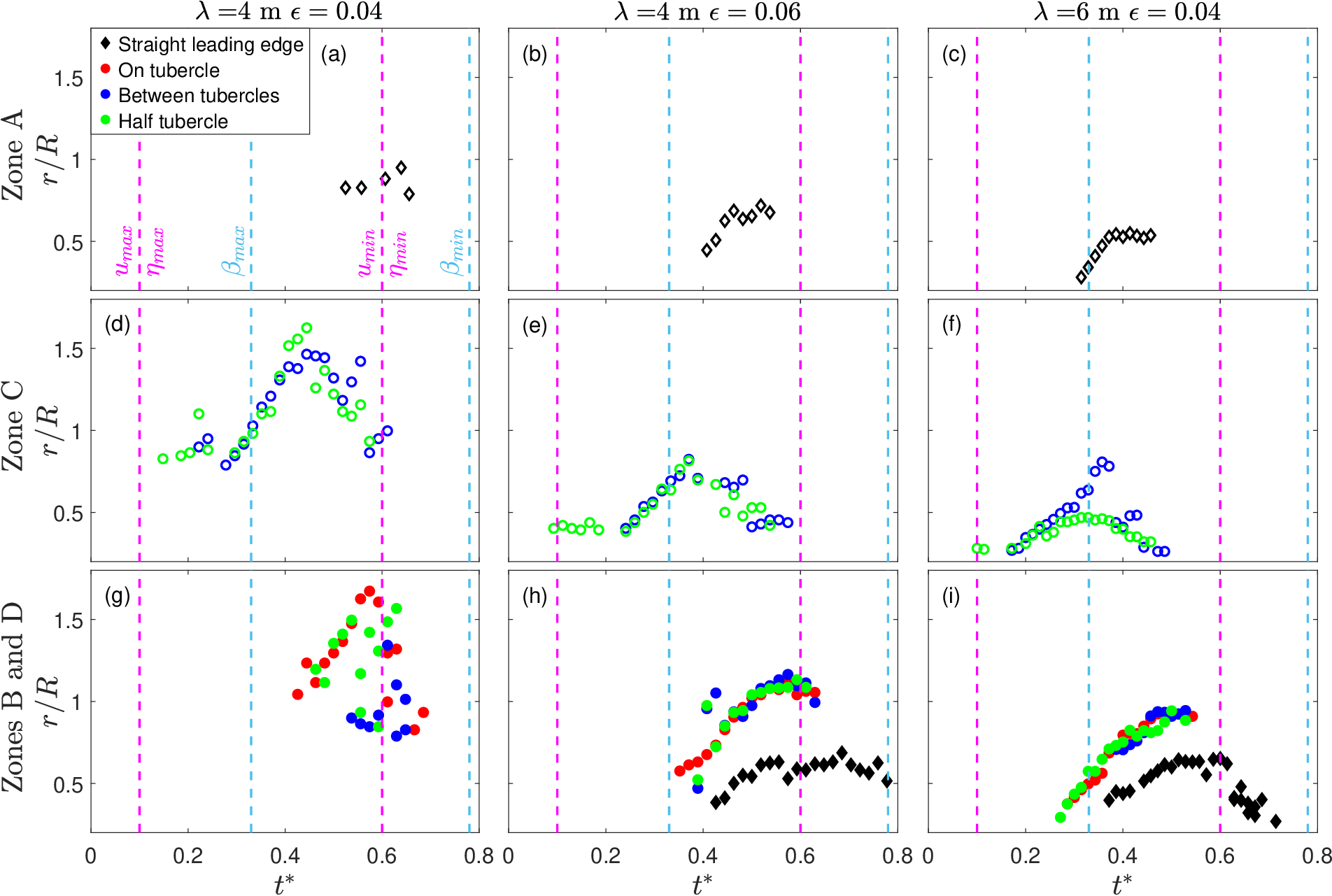}
    \caption{Evolution of the non-dimensional effective radius, $r/R$, of the different vortex structures as a function of the non-dimensional time $t^*$ for the three wave forcing cases: (a,d,g) $\lambda=4$~m, $\epsilon=0.04$, (b,e,h) $\lambda=4$~m, $\epsilon=0.06$, and (c,f,i) $\lambda=6$~m,$\epsilon=0.04$. The different vortices are represented separately: (a,b,c) zone A, (d,e,f) zone C and (g,h,i) zones B and D. Symbol shape and colour indicate different planes of measurement, i.e., straight leading edge (black diamonds), measurement plane coincident with a tubercle peak (red dots), measurement plane coincident with a tubercle trough (blue dots), and measurement plane halfway between tubercle peak and trough (green dots). Vertical dashed magenta lines refer to the instants of maximal and minimal horizontal velocity $u_x$ (wave peak and trough) and the blue vertical dashed lines to the instants of maximal and minimal angle of attack $\beta$.}
    \label{Radius_vortex}
\end{figure}

In order to fully characterise the vortices' formation, we also consider their spatial dimension over time. We define vortex size by computing an effective radius $r$, defined as $r=\sqrt{\mathcal{A}/\pi}$, where $\mathcal{A}$ is the measured area in which $Q>100$. 
This radius can be compared to $R$, the theoretical typical length scale of the wave-induced flow, which is defined as the vertical displacement

\begin{equation}
    \phantom{,}R=\mathcal{R}\tan\alpha_m,
\end{equation}
made by a fluid particle over a half period of the wave forcing, at the hydrofoil nominal immersion depth, $z=-h$, 
with $\mathcal{R}=Ae^{-\frac{2\pi h}{\lambda}}$ and $\tan(\alpha_m)=\frac{\mathcal{R}\omega}{u_0}$ given by the amplitude of variation of the angle $\alpha$ from Eq.~\eqref{tanalph_eq}. The variation of $r/R$ with non-dimensional time $t^*$ is plotted in Fig.~\ref{Radius_vortex} for the different vortex structures, (a,b,c) zone A, (d,e,f) zone C, and (g,h,i) zones B and D and for the different wave forcing conditions. Data are given for each vortex up to the moment that the vortex begins to leave the measurement domain. For the straight-leading-edge hydrofoil (black diamonds), the two vortices (boundary layer roll up at trailing edge, zone A, and leading-edge vortex, zone B) quickly grow until approximately equivalent radii of $r\approx0.6R$. For the tubercled hydrofoil, the leading-edge region of vorticity accumulation (zone C) first grows in spatial extent, before shrinking from $t^*\approx0.4$ both between tubercles and in the half-tubercle measurement plane (blue and green, respectively). The large leading-edge vortex (zone D) emerges subsequently in every measurement plane at the same time. For $\lambda=4$~m and $\epsilon=0.04$ [Fig.~\ref{Radius_vortex}(g)], the vortex radius increases over time until $r/R\approx1.5$, with experimental data being relatively spread. For the two other cases (b-c), the large diffused leading-edge vortex grows  
from $t^*\approx0.35$ to $t^*\approx0.5$ in every measurement plane. Its radius grows until $r\approx R$ and tends to maintain this value until the vortex leaves the measurement domain. As seen in Fig.~\ref{XZ_vortex}(b,c), in the tubercled cases, the leading-edge vortex is advected almost horizontally, without following the hydrofoil surface. 
The tendency of the vortex effective radius to converge to $r=R$ supports an interpretation that the spatial extent of leading-edge vortex formation is linked to the vertical displacement length scale of the wave induced fluctuating flow. This length scale appears to play a more direct role in the case of the tubercled leading edge, where the wave-induced vertical velocity component can penetrate between tubercles. 

\begin{figure}[h!]
    \centering
    \includegraphics[width=0.6\linewidth]{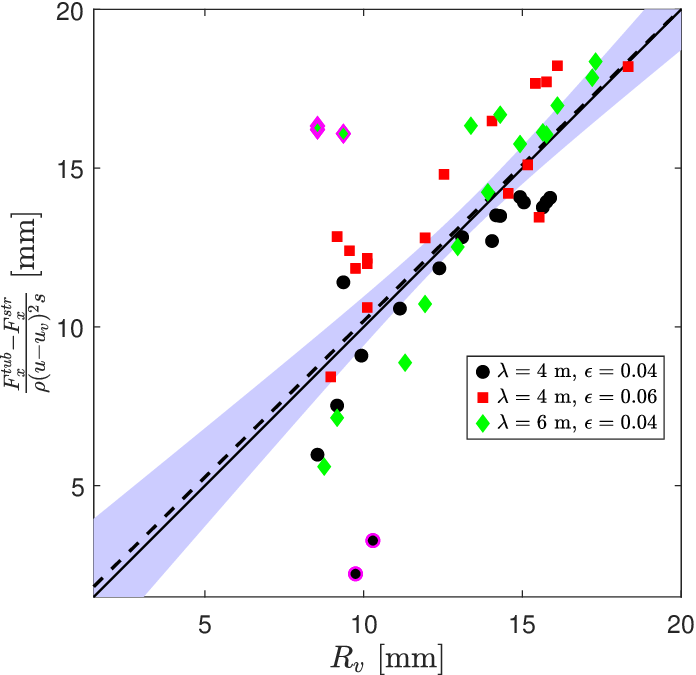}
    \caption{Effective length scale of the leading-edge vorticity accumulation region, as predicted by
    the difference in observed horizontal force between straight leading edge and tubercles hydrofoils, following Eq.~\eqref{Fxdiff_model} and assuming $C_{dv}=1$. 
    The three stall cases are represented here as: $\lambda=4$~m, $\epsilon=0.04$ (black dots), $\lambda=4$~m, $\epsilon=0.06$ (red squares), and $\lambda=6$~m, $\epsilon=0.04$ (green diamonds). The black dashed line is a best linear fit to the experimental data and yields to a slope of $0.98\pm0.18$ and vertical-axis intercept of $0.35\pm2.39$, with R-square of 0.74. The clear outliers outlined in magenta are not taken into account in the fit. Blue shading indicates the 95$\%$ confidence interval of the fit. The black solid line (slope 1) is the prediction of Eq.~\eqref{Fxdiff_model}.}
    \label{FxTub-FxNoTub_Rvort}
\end{figure}

We can now use our observations of the flow behaviour to mechanistically explain the effect of tubercles on the generation of horizontal force $F_x$, which is plotted in Fig.~\ref{etaFxFzNotubTub} and \ref{FxTub-FxNoTub_t}. The observed difference in $F_x$ occurs primarily in the time interval $t^*\in[0.2,0.6]$, which corresponds to conditions where the angle of attack is maximum and the vortex structures on the suction side of the hydrofoil are in their early stages of formation. 
As observed in Fig.~\ref{etaFxFzNotubTub}, the effect of tubercles on the value of $F_x$ in the steady case (without waves) is small compared to the magnitude of the force fluctuations in the unsteady cases. 
We, therefore, for expediency in modelling the effect of tubercles, assume that the steady-state drag coefficient is approximately identical for the two leading-edge geometries.  
We hypothesize the same for added mass, meaning that the observed force difference between straight- and tubercled-leading-edge cases can primarily be attributed to modified stall behaviour in the presence of tubercles. The primary qualitative flow difference between the straight-leading-edge and tubercled cases investigated here is an accumulation of vorticity near the leading edge, within the first third of the chord length. 
This accumulation relates to the separated shear layer which will later form the leading-edge vortex; 
however, in the tubercled case, it emerges earlier within each wave period, remains distinct from the leading-edge vortex as indicated by the $Q$-criterion, and exhibits minimal downstream advection. 
The downstream advection rate $u_v$ observed is only 
between $0.04u_x$ and $0.11u_x$ in $t^*\in[0.2,0.6]$ [Fig.~\ref{XZ_vortez_t}(a-c)]. The motion of this vortex structure with the hydrofoil, near its leading edge, presents an additional obstruction to the horizontal passage of the external flow. We evaluate the consequence for the drag force if we make the simplistic assumption that this structure's effect can be considered analogous to that of a bluff body of equivalent dimension, given by its effective radius 
$R_v=r(x/c<1/3)$, moving at the observed advective velocity $u_v$.
To model this assumption, we introduce an additional steady-state drag term, 

\begin{equation}
     \phantom{,}F_x^{tub}-F_x^{str}=C_{dv}R_v \rho(u-u_v)^2s,
\label{Fxdiff_model}
\end{equation}
with $C_{dv}=1$, corresponding to the value for a circular cylinder~\citep{heddleson1957summary}. If these assumptions are sufficiently accurate to explain the observed horizontal force difference between straight-leading-edge and tubercled cases, 
the constructed length scale $F_x^{tub}-F_x^{str}/[\rho(u-u_v)^2s]$ should be equal to the observed effective radius $R_v$ of the accumulation of leading-edge vorticity. 
Fig.~\ref{FxTub-FxNoTub_Rvort} shows the relationship between the experimentally determined values of these two variables. 
A linear regression on the data for the all three cases, $\lambda=4$~m, $\epsilon=0.04$ (black dots), $\lambda=4$~m, $\epsilon=0.06$ (red squares), and $\lambda=6$~m, $\epsilon=0.04$ (green diamonds) yields a slope of $0.98\pm0.18$ and vertical-axis intercept of $0.35\pm2.39$, with R-square of 0.74 (dashed line). The solid line of slope 1 represents the prediction of Eq.~\eqref{Fxdiff_model}, and lies well within the 95\% confidence interval (blue shading) of the linear regression. 
The few clear outliers (markers with magenta borders) originate from sensitivity of the $Q=100$ contour 
to experimental measurement noise, which in occasional cases can artificially displace the estimated vortex cent by a large amount (see $movie5$, $movie7$ and $movie8$). These outliers are not taken into account in the linear fit. 
Including these outliers in the linear regression yields a similar slope of $0.89\pm0.26$, intercept of $1.73\pm3.47$, with R-square of 0.48. The prediction of Eq.~\eqref{Fxdiff_model} remains in that case well within the 95$\%$ confidence interval of the fit. This surprisingly simplistic model explains much of the variation in force 
caused by stall in the presence of large amplitude wave forcing, using only knowledge of the dynamic stall vortex dimensions and velocity. 
Force variation in unstalled conditions, 
$\lambda=2$~m in Fig.~\ref{FxTub-FxNoTub_t}(a) for example, cannot of course be accounted for by this method. 
Further study is needed to better understand the bounds of the assumptions made in this work, and extend it across a greater range of forcing conditions and stall regimes. 
Nevertheless, the present work quantifies for the first time with direct force measurements how tubercles modify horizontal forces on a foil undergoing dynamic stall. Furthermore, we propose a simple model, which explains to a first approximation tubercles' effect on horizontal force generation by reference to experimentally observed flow behaviour.

\section{Conclusion}\label{conclusion_section}
In this study, we have described the hydrodynamic interaction of a horizontally oriented hydrofoil with the flow field imposed by gravity waves, through which it was traversed at a constant speed. The focus was on wave forcing conditions spanning regimes of unsteady attached flow, weak quasi-static stall phenomena, and dynamic stall with leading-edge boundary-layer separation. Both force generation and the flow field over the hydrofoil suction surface were quantified. 

A hydrofoil with a straight leading edge was first considered. A comparison was then made with an equivalently dimensioned hydrofoil fitted with leading-edge tubercles: A passive flow-control technique, which has been widely studied under steady flow conditions, but for which there exists a dearth of information regarding their effect in unsteady flow. In the present work, we have extended the extant data on the operating principles and effects of leading-edge tubercles in unsteady flow by examining their interaction with the flow induced by wave forcing; a flow which exhibits multi-component unsteadiness, and one which has obvious relevance to marine applications in naval engineering, offshore flow energy harvesting, and even with regard to understanding the hydrodynamics of the humpback whale, \textit{M. novaeangliae}, which inspired the tubercle geometry. 

The data used in this investigation were generated by experiments performed in a 142 meter long towing tank equipped with highly resolved measurement systems, including force transducers and a submerged, torpedo-mounted PIV system. 
The horizontal and vertical forces acting on the hydrofoil were measured, while PIV-based flow observation and quantification was simultaneously performed. This measurement pairing, which has never, to the best of our knowledge, previously been performed in describing dynamic stall generated by gravity waves, enabled us to link the observed force generations mechanistically with unsteady flow features. 

We observed that the waves impose temporal variation of the local flow orientation and speed at the location of the hydrofoil. The interaction of this fluctuating flow with the hydrofoil induces forces acting on the hydrofoil. Adding steady-state lift and drag estimates based on instantaneous flow velocity and angle of attack to a potential-derived added-mass component proves inadequate to capture the observed force variations at large wave forcing amplitudes.  
It is under these large amplitude wave forcing conditions that dynamic stall is triggered. 
In the dynamic stall regime, an accumulation of vorticity near the leading edge develops as the hydrofoil's angle of attack is dynamically increased. This vorticity accumulation intensifies until the flow near the hydrofoil surface reverses and a classical leading-edge vortex is formed. Since our experiments span a range of wave forcing amplitudes, we also observed intermediate cases between unsteady attached flow and dynamic stall. 
We refer to these intermediate cases, and their related flow phenomena as ``quasi-static'' stall, due to the low dimensionless pitch rates at which these phenomena occur. The primary marker of the quasi-static stall regime is an initial boundary-layer separation location close to the trailing edge, which progressively migrates forward. If leading-edge separation occurs during this separation line migration, the resulting leading-edge vortex can perturb the separated flow near the trailing edge. In our case, this provokes shear-layer roll-up and the formation of a second vortex on the suction surface, downstream of the primary leading-edge vortex structure.

By applying a spanwise-sinusoidal tubercle geometry to the leading edge, the evolution of the separated flow structures was modified: The initial leading-edge accumulation of vorticity emerged first in the troughs between tubercles. This accumulation appeared earlier in our measurements in the tubercled case than in the case with a straight leading edge. The early presence of this vorticity accumulation appears to prevent trailing-edge separation from occurring. 
The consequence of this is that the circulation, which in the straight-leading-edge case is distributed across two vortices, is in the tubercled-leading-edge case concentrated in a single, larger, leading-edge vortex. Interestingly, despite the spanwise varying leading-edge geometry, the leading-edge vortex location appears largely spanwise invariant.

The leading-edge vortex remains attached to the leading edge via a shear layer, as is typical in dynamic stall flows, for a significant portion of the wave period. Notably, however, the near-leading-edge vorticity accumulation, at the origin of this shear layer, remains in place, displaying minimal downstream advection. In this vortical region, the $Q$-criterion is positive in a region distinct from the primary leading-edge vortex, indicating a localized dominance of rotation over shear. Here, we have thus considered this region as a distinct vortex structure, whose spatial characteristics can be used to predict the modified force generation due to the presence of tubercles. 

The two extant studies on the dynamic stall of foils with leading-edge tubercles (\citet{hrynuk2020effects} and \citet{valls2025dynamic}) reported an effect of tubercles on the leading-edge vortex circulation, and supposed that this would also translate to an effect on lift. Only \citet{valls2025dynamic} supported this with a numerical computation of the lift of a pitching tubercled foil. Our direct experimental force measurements indicate a different behaviour: The leading-edge vortex is strengthened by the presence of tubercles, but lift generation is not impacted to a practically relevant degree. 
The negligible lift generation effect follows from Kelvin's circulation theorem and the application of the classic Kutta-Joukowski lift theorem.  
Here, we have also provided the first experimental quantification of the effect of tubercles on drag force production, which was not considered in the two previous dynamic stall studies~\citep{hrynuk2020effects,valls2025dynamic}. A simple quasi-steady linear model for the force generation indicates that, under wave forcing, the horizontal force on a hydrofoil is dominated by added-mass effects, leading to periodic fluctuations in the horizontal force of sufficient magnitude as to even provide negative drag at certain periods of the wave forcing. When tubercles are introduced, these negative drag peaks are attenuated, effectively reducing the magnitude of the temporal drag force fluctuations. The introduction of tubercles thus has clear potential to practically benefit any submerged structure or turbine blade by diminishing fatigue loading due to waves, at the expense of an increased average drag.

The phase of wave forcing in which increased drag was observed on the tubercled hydrofoil corresponds to the period in which the dynamic stall vortex develops. Observing that the vorticity accumulation near the leading edge barely advects downstream, we modelled its effect on drag via analogy to the effect of a bluff body of comparable dimension. This rather simplistic model was surprisingly good at predicting the drag increase due to tubercle-induced flow field modification, thus establishing a mechanistic explanation for the observed drag increase based on the observed unsteadily separating flow behaviour. 

The present work provides, hence, a first observation and quantification of hydrofoil interaction with gravity waves, while demonstrating the passive manipulation of the flow and forces arising from this interaction. 
Future work will turn attention to spanwise flow effects and extend to the more complex case of hydrofoil rotation within the wave field, such as is relevant to hydrokinetic turbine applications. 

\vspace{1em}

\noindent \small \textbf{Acknowledgements} {The authors thank J.\ den Ouden, S.\ Tokgoz and P.\ Poot for technical assistance in preparing and performing the experimental measurement campaign.}
\vspace{1em}

\noindent \small \textbf{Funding} {A-J.\ Buchner was supported by the Netherlands Organisation for Scientific Research (NWO), under VENI project number 18176.}

\bibliographystyle{jfm}
\bibliography{jfm}

\end{document}